\documentclass[12pt]{article}
\usepackage[a4paper, margin=0.82in]{geometry}
\usepackage{hyperref}
\usepackage{ellipsis}
\usepackage{mathrsfs}
\usepackage{color}
\usepackage{psfrag}
\usepackage{graphicx}
\usepackage{caption,subcaption}
\usepackage{enumerate}
\usepackage{amsmath,amssymb,calc,amsfonts}
\usepackage{latexsym}
\def\beq{\begin{eqnarray}}
\def\eeq{\end{eqnarray}}

\def\={\stackrel{\Delta}{=}}

\usepackage[utf8]{inputenc}
\title{Gravitational collapse of matter fields in de Sitter spacetimes}
\author{Akriti Garg \footnote{Email: akritigarg571@gmail.com} and 
Ayan Chatterjee \footnote{Email: ayan.theory@gmail.com}\\
Department of Physics \& Astronomical Science, \\Central University of Himachal Pradesh, Dharamshala-
176215, India.}
\usepackage{graphicx}
\begin{document}
\date{}
\maketitle

\pagenumbering{arabic}
\renewcommand{\thesection}{\arabic{section}}
\begin{abstract}
In this paper, we discuss the spherically symmetric gravitational collapse of matter fields in the de Sitter universe. The energy- momentum tensor of the matter field is assumed to admit a wide variety including dust, perfect fluids with equations of state, fluids with tangential and radial pressure, and with bulk and shear viscosity. Under different initial conditions imposed on the velocity and the density profiles, and by combining the results from exact analytical methods with those obtained from numerical techniques, we track the formation and evolution of spherical marginally trapped spheres as the matter suffers continual gravitational collapse. We show that the quasilocal formalism of trapped surfaces
provides an ideal framework to study the evolution of horizons. More precisely, black hole and cosmological horizons may be viewed as the time development of marginally trapped surfaces. 
\end{abstract}

\maketitle



\section{Introduction}
The complete understanding of gravitational collapse of matter fields remains one of the most challenging objectives of gravitational physics \cite{Hawking_Ellis,  Wald, Landau_Lifshitz, joshi}. Although tremendous progress have already been made, most particularly 
in the developing mechanisms to comprehend the formation of spacetime singularities and horizons, the examples remain mostly confined to spherically symmetric models, since a systematic progress in analytical study of collapse of even an axisymmetric configuration of matter fields is yet to be grasped in its full glory (see \cite{Abrahams:1994nw}, and a recent study in \cite{Rostworowski:2025hvj}). In the absence of such developments, it remains to strengthen our present understanding by allowing more generality. For example, a lingering issue associated with the spherical collapse of matter fields, leading to formation of black holes and a central spacetime singularity, has been concerning the peculiarities regarding initial conditions on the gravitational and matter degrees of freedom which allow for the continuous gravitational collapse to take place \cite{Penrose:1964wq, Penrose:1969pc}. Indeed, Penrose's conjecture of (weak) cosmic censorship only provides a vague guide, stating in effect that, under the assumptions of correctness of the Einstein theory of gravity, and \emph{reasonable} conditions on the matter equations of state and energy- momentum tensor, \emph{generic} regular initial conditions on a  \emph{suitable} Cauchy hypersurface lead to spacetime singularity which remains causally disconnected from asymptotic observers \cite{Hawking:2010nzr, Penrose:1999vj}. This is the presently accepted picture of a black hole horizon. The (strong) cosmic censor leaves out initial singularities like the big bang since it prohibits the existence of locally naked singularities \cite{Hawking:2010nzr, Wald}. However, the exact requirements under which the either versions of the cosmic censor may be deemed reasonable is debated in the literature (recent reviews are in \cite{Landsman:2021mjt, Landsman:2022hrn, Joshi:2011rlc}). So, in this paper, we shall look into the issue of gravitational collapse of matter with \emph{reasonable} requirements on the initial conditions of velocity and density and shall determine the role of the positive cosmological constant in gravitational collapse.  \\

The study of gravitational collapse of matter fields leading to black hole formation has a long history, with the first study being carried out by Dutt \cite{SD}, and by Oppenheimer and Snyder (OSD) \cite{OS}, on a spherical configuration of homogeneous dust. The OSD model is simple but furnishes important conclusions: First, the spherical ball of matter, in finite proper time, collapses past its Schwarzschild radius to reach the central singularity of infinite density. Second, just as the matter crosses its Schwarzschild radius, no light is able to escape to reach observers at the asymptotic infinity. This simple construction is the basis for understanding formation of compact objects like black holes in the universe. However, it is argued that real physical configurations are not homogeneous and therefore, the OSD formulation must receive modifications to conform to physically acceptable situations. The Lemaitre- Tolman- Bondi (LTB) model \cite{Lemaitre:1933gd, Tolman:1934za, Bondi:1947fta, Misner_Sharp, Hernandez:1966zia} incorporates inhomogeneities in the collapsing matter configuration, and is therefore considered as a model better suited to analyze gravitational collapse. Further considerations on these models have also been carried out in the last several decades, and a comprehensive guide may be obtained in \cite{Stephani:2003tm}. In the issue related to gravitational collapse in the deSitter spacetimes, the pioneering study was carried out in \cite{Omer}, while later studies have made significant improvements by incorporating other kinds of matter fields (see \cite{joshi, Krasinski:1997yxj, Plebanski:2006sd, Lake:2000rm, Markovic:1999di, Deshingkar:2000hd, Cissoko:1998mx, Debnath:2006nh}
for specific examples of dust models). However, a detail study of viscous fluids, perfect fluids with equations of state, and fluids with tangential pressure only have not been carried out, and the goal of the present paper is to incorporate them into the formalism, and to determine the extent to which these fluid features may affect the dynamics
of gravitational collapse. In particular, we would like to ascertain the scale to which the horizon and singularity formation time may be altered significantly
due to the presence of tangential pressures, or the shear or bulk viscosities. 
It is expected that this study shall also shed some light into the issue of formation of naked singularities in the Einstein theory (various aspects of naked singularities, including their existence, strength of singularity and related issues are detailed in \cite{Eardley:1978tr, 
Christodoulou_1984, Papapetrou, Newman:1985gt, Ori:1987hg, Newman_joshi, 
Dwivedi:1989pt,Ori:1989ps,
Lake:1991bff, Lake:1991qrk, Joshi:1993zg, Christodoulou_1994, Dwivedi:1994qs,Singh:1994tb,
Jhingan:1996jb,Harada:1998cq,Harada:1998wb,Maeda:1998hs,Joshi:1998su,Maeda:2006pm,Goswami:2006ph, Hassannejad:2023lrp, Shojai:2022pdq}). \\  

%
%
\begin{figure}[htb]
\begin{subfigure}{.23\textwidth}
    \centering
    \includegraphics[width=4.3 cm]{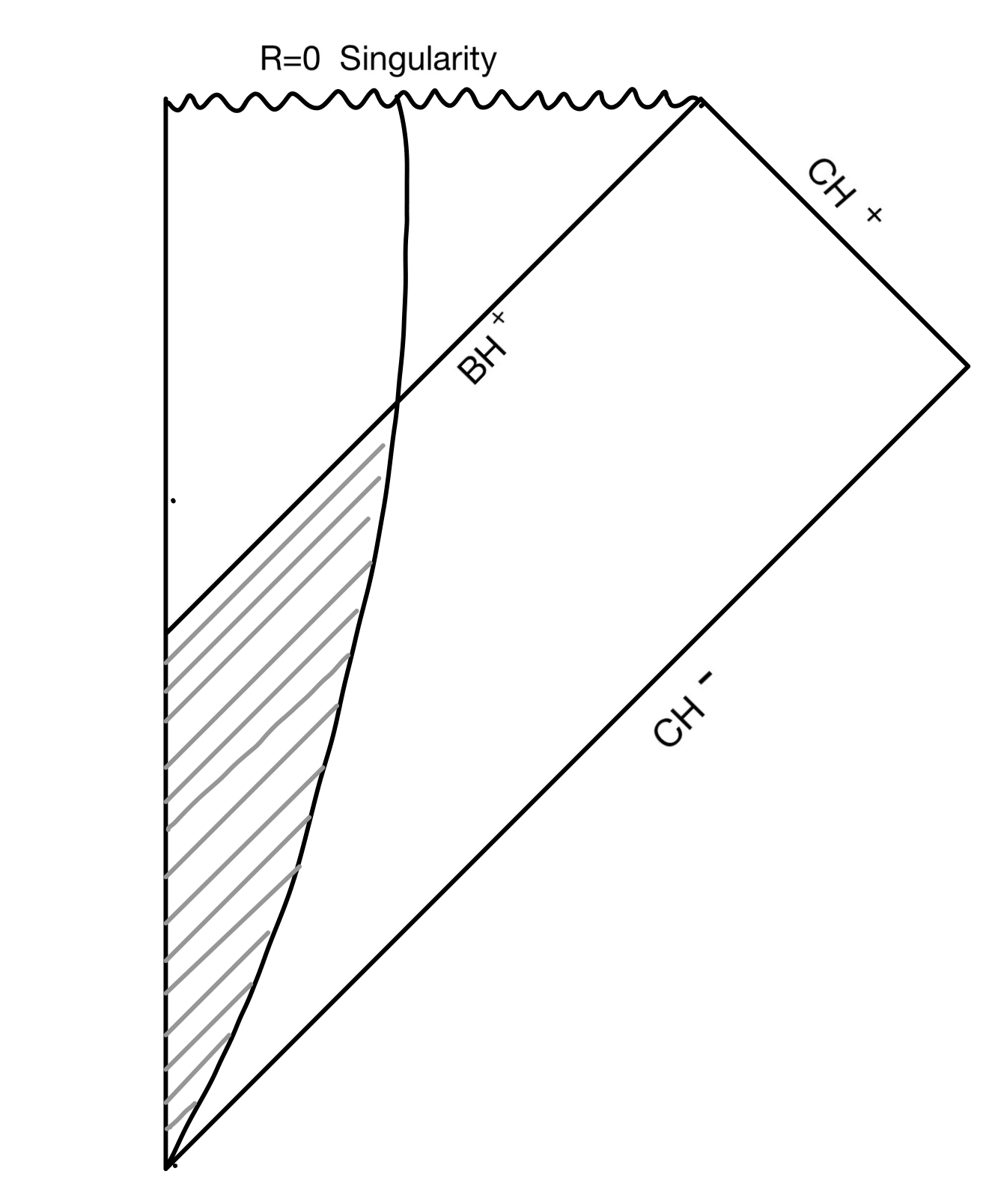}
    \caption{}
    \end{subfigure} %
    \qquad\qquad 
    \begin{subfigure}{.43\textwidth}
    \centering
    \includegraphics[width=9.8cm]{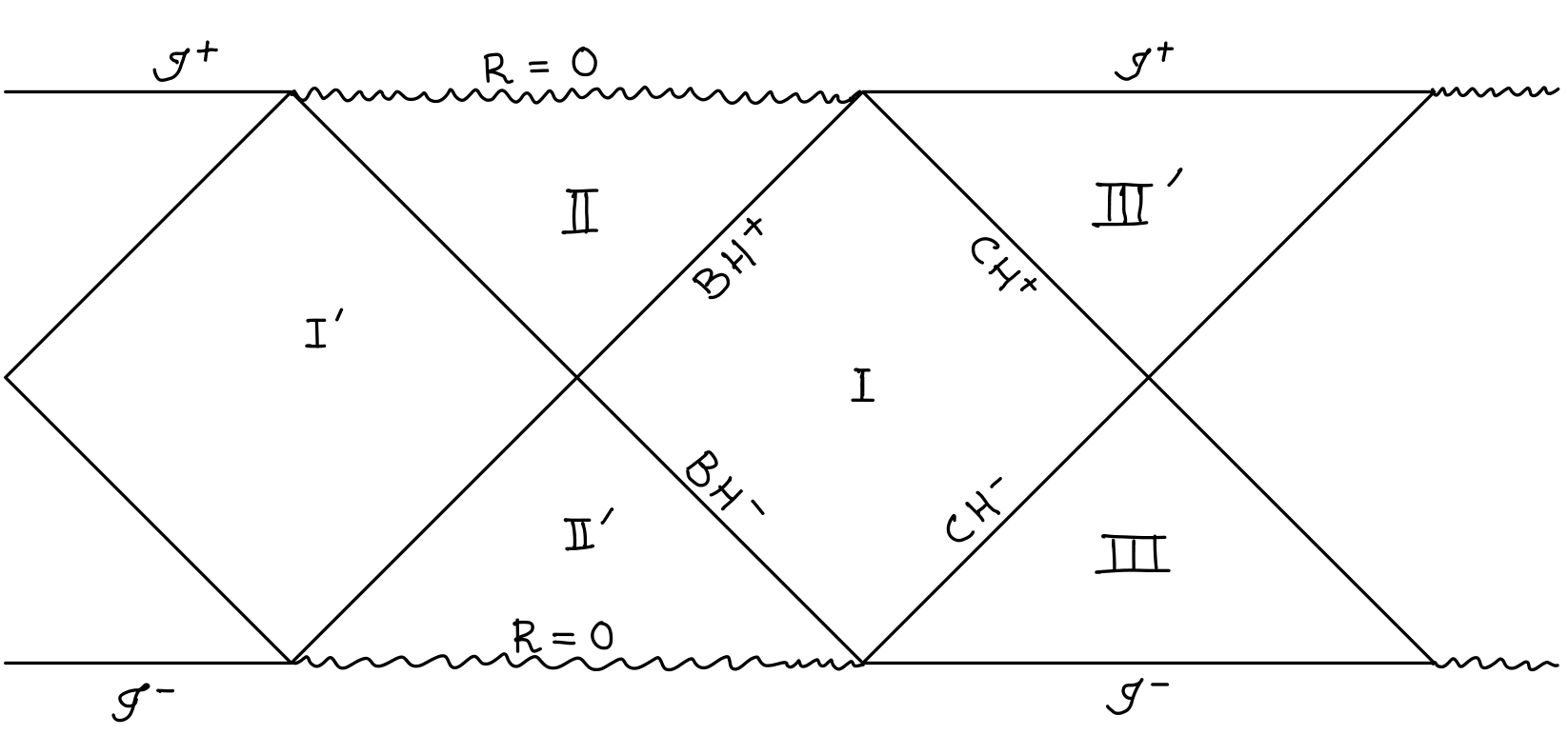} %
    \caption{}
    \end{subfigure} %
 \caption{The figure (a) is the Penrose diagram describing the formation of black hole due to collapse of matter in a de Sitter spacetime. The $CH^{-}$ and $CH^{+}$ describes the past and the future cosmological horizons, whereas $BH^{+}$ is the future black hole horizon. The figure (b) is the extended Penrose diagram of the Kottler spacetime. In the region $I$, the spacetime is static and the observer may move in increasing or decreasing directions of $R$. The regions $II$ or $III$ are such that $(\partial/\partial t)$ is spacelike. Signals may be sent from $I$ to $II$, and from $III$ to the static region. }
 \label{fig_BH}%
\end{figure}
Before we go on to the study of gravitational collapse formalism, let us discuss black hole horizons in deSitter spacetimes. Black hole solutions, both rotating and non- rotating, in the deSitter spacetimes are well known. Out of these, the Kottler metric is the unique spherically symmetric solution to the vacuum  Einstein field equation with a cosmological constant, $R_{\mu\nu}+\Lambda \, g_{\mu\nu}=0$ \cite{Kottler}. It is a generalisation of the Schwarzschild metric with a cosmological constant, and therefore, it is often referred to as the Schwarzsxhild- de Sitter solution when $\Lambda >0$. The Birkhoff theorem also holds here, and this metric is the unique spherically symmetric static metric, and is given by:
\begin{equation}\label{Kottler_metric}
ds^{2}=-f(r)\, dt^{2} +f(r)^{-1}\, dr^{2}+r^{2}\, d\Omega_{2}\, 
\end{equation}
where $f(r)=\left[1-(2M/r)-(\Lambda/3)\, r^{2}\right]$. The zeros of this function $f(r)$ indicates 
the presence of horizons. This required function is a third order polynomial equation $\mathcal{H}(r)=(\Lambda/3)r^{3}-r- 2M$, with derivative (with respect to $r$ given by $\mathcal{H}^{\prime}(r)=(\Lambda r^{2}-1)$. The function $\mathcal{H}(r)$, for the choice of $\Lambda>0$, must have a minimum for $r\in [\,0, \infty \,)$, such that 
the polynomial itself has either no zero, two degenerate zeros, or two positive and real zeros. The degenerate zeros occur at $M\sqrt{\Lambda_{d}}=2/3$, for which the horizons are located at $r=3M$. For $\Lambda \in (0, \Lambda_{d})$, the horizons, called the black hole horizon $r_{BH}$
and the cosmological horizon $r_{CH}$ are such that $r_{CH}> r_{BH} >2M$ (the precise values of these horizons shall be discussed in the section 3 of the paper). Naturally, for the degenerate solution, 
the horizons merge and the resulting metric is called the Nariai solution \cite{Nariai}. The Penrose diagram for the Kottler spacetime and for the collapsing spacetime is described in Fig. \eqref{fig_BH}. The black hole solution in eqn. \eqref{Kottler_metric} is asymptotically deSitter, and although the value of $\Lambda$ of our present universe in 
quite small
so as to be practically negligible at large distances from the massive source, its mere presence creates 
difficulties in understanding the subtleties of gravitational collapse formalism. Therefore, for the purpose of gaining conceptual clarity about the theoretical framework of formation and evolution of horizons, particularly in presence of fluids with involved equations of state, it is important that the formulation of gravitational collapse is understood for deSitter spacetimes. \\

The other issue, which shall be one of the main themes of this paper, is related to the quasilocal formulation of horizons and their suitablility to clearly manifest the nature and classical geometrical properties of horizons during their time development \cite{Dreyer:2002mx, Ashtekar:2004cn, Booth:2005qc, Schnetter:2006yt, Booth:2005ng, Booth:2006bn, Shapiro:1983du, Baumgarte:2010ndz}. To pinpoint the matter in question, it needs to be asserted that the global descriptions of horizon like the Event Horizon (EH) is teleological, and is therefore its applicability to physically acceptable situations, and even in numerical simulations, is open to question (see \cite{Hayward:1993wb, Chatterjee:2012um, Ashtekar:2000sz, Ashtekar:2000hw, Ashtekar:2002ag, Ashtekar:2003hk, Ashtekar:1997yu, Chatterjee:2014jda, Chatterjee:2015fsa} and the recent review \cite{Ashtekar:2025wnu} 
for a detailed application of the quasilocal formalism to classical, quantum and numerical relativity). More precisely, the location of the EH (consider for example, a black hole in asymptotically flat spacetime) may only be specified if the entire asymptotic null infinity is known, since EH is defined as $\partial[J^{-}(\mathscr{I}^{+})]$ \cite{Hawking_Ellis, Wald}. In situations where the black hole spacetime is in equilibrium, that is, a timelike Killing vector field exists in the entire spacetime, the $\mathscr{I}^{+}$ is precisely known and therefore, the EH is a true description of the null boundary of black hole region. For the purpose of black hole mechanics, this horizon also admits a null Killing vector field signifying that the horizon is in equilibrium, and that no matter field is allowed to cross the black hole boundary \cite{Hawking_Ellis, Wald}. In short, the location of the black hole (or for that matter, the cosmological horizon too), in the EH description may be stated as problem where future affects the past, and therefore, is not an ideal set-up to study dynamics of physically interesting systems (since for these teleological systems, the dynamics at the present needs specification of future boundary conditions on the variables involved). In particular, for numerical simulations of black hole evolution, the EH is not an ideal framework.  To solve such problems, the quasilocal description of black holes, called marginally trapped tube (MTT) has been developed \cite{Booth:2005ng, Ashtekar:2005ez, Andersson:2005gq, Andersson:2007fh, Booth:2010eu, Bengtsson:2008jr, Bengtsson:2010tj, Bengtsson:2013hla, Booth:2012rm, Booth:2017fob, Schnetter:2005ea, Nielsen:2010wq}, which in turn, is based on the formulation of trapped surface introduced by Penrose to study black hole interior and the singularity theorems. Before going to the description of MTT, it is worthwhile to recall the idea of trapped surfaces. A trapped surface $\mathbb{S}^{2}$ in a $4$- dimensional spacetime is a $2$- dimensional closed spacelike surfaces such that the expansion $\theta$ of the two future directed orthogonal null vector fields $\ell^{a}$ (outgoing) and $n^{a}$ (ingoing) to $\mathbb{S}^{2}$ are both negative. In case the expansion for $\ell^{a}$, given by $\theta_{(\ell)}$, vanishes while that for the ingoing vector field, $\theta_{(n)}$, is negative, the surface is called \emph{marginally trapped}. Trapped surfaces (and their marginal versions) are of immense importance for black hole physics since the singularity theorems asserts that if a trapped surface is formed, a spacetime singularity is inevitable in the future \cite{Hawking_Ellis}. For this reason the search for trapped surfaces in the numerical simulation of spacetime is of primary interest, and several tools have been developed to carry out such checks. In the MTT formalism, the horizon is built up by stacking marginally trapped surfaces
in the form of a cylinder such that $\mbox{MTT}=\mathbb{S}^{2}\times \mathbb{R}$, where $\mathbb{R}$ indicates the future direction of evolution. The MTT, in contrast to the EH, is not a null surface- it may admit all the three signatures depending on the physical situation: if matter field is falling through the MTT, it is spacelike, whereas if matter fields exit, MTT is timelike. The MTT is null if no matter crosses it and is said to be in equilibrium. The MTTs which are spacelike or timelike are also called Dynamical Horizons (DH), while the MTT in equilibrium is also referred to as the Isolated Horizon (IH) (\cite{Ashtekar:2000sz, Ashtekar:2000hw, Ashtekar:2002ag, Ashtekar:2003hk, Ashtekar:2025wnu}). The situation becomes interesting in the presence of
perfect fluids and viscous fluids admitting bulk and shear viscosity. In each of these circumstances we shall use the field equations to evolve the initial density profile in time so as to locate the spherical MTT, and then illustrate its ability to faithfully depict the expected evolution of horizon: when matter falls, the black hole horizon grows (and is spacelike), the cosmological deSitter horizon decreases (has a timelike signature), and both horizons remain null when no matter cross eith horizon. In each case, as we shall show, the MTT is spectacularly successful as a description of horizon, in equilibrium or otherwise.\\

The other question that the present paper deals with, is related to the possible wide variety in the energy-momentum tensor that fluids may admit, and how the dynamics of gravitational collapse may be affected
due to such qualities. In particular, as we have alluded to above, we
shall consider fluids with tangential and radial pressures, perfect fluids, as well as fluids with bulk and shear viscosity. What are the expectations due to inclusion of matter of such types? We expect that the process of gravitational collapse must slow down since viscous forces usually generate anisotropies, which in turn leads to a more sluggish evolution. To see if the expectation is in the affirmative, we have included a section on the the study of gravitational collapse of dust matter in deSitter spacetimes so that a comparison may be made. Although some studies in this direction have been carried out earlier, we solve the equations of motion using exact analytical methods in detail and then, use them to track the evolution of spherical marginally trapped surfaces \emph{vis-a-vis} the fall-in of shells of the matter configuration towards the central singularity. We show that, as the spherical ball of matter collapses, shell by shell, towards the central singularity, the spherical MTT evolves and matches with the IH just the matter density vanishes. The situation remains identical due to collapse of fluids with pressure or with viscosity. But what alters is the time that the shells take to reach the central singularity or, the horizon formation time. The effect of viscosity is to increase these times by a few orders of magnitude, or in other words, make the process of collapse slower, as was expected. Even the time taken to reach the Nariai limit (when the black hole and the cosmological horizons merge), is increased manifold. \\

The paper is presented as follows: In the next section \ref{sec2}, we introduce the basic notation and mathematical formalism to develop the theory of gravitational collapse of matter in deSitter spacetimes. This section shall include the metric, the geometrical variables associated with it, the field equations, fluid properties of the collapsing matter, as well as a detailed discussion on the regularity requirements on the matter and geometrical variables. 
These regularity conditions shall have crucial use in the next sections.
In the section \ref{section3}, we use the formalism for dust matter which are either marginally bound or gravitationally bound on the initial data surface. We determine the time required for the collapsing matter to reach singularity, and determine how various initial density profiles affect the evolution of quasilocal black hole and cosmological horizons (MTT). This study shows the power of MTT as it is able to deal with both the future and the past horizons within the same structure with equal ease. The next section, Section \ref{section4} repeats the analysis of the previous section for fluids of various types, and in each case, obtains the spacetime singularity, the horizon formation time and the relative growth of the black hole and cosmological MTTs. A discussion of the results is carried out in the last section. The Appendixes contain material pertaining to the various calculations in the text, and may be referred for detailed explanations.\\  


\section{Spherically Symmetric Collapse Formalism} \label{sec2}

In the following, we begin our study of gravitational collapse of matter fields leading to the formation of trapped surfaces and black hole horizon in a deSitter spacetime. Let us consider a general spherically symmetric metric expressed in the following form:
\begin{equation}\label{guv}
ds^2 = -e^{2 \alpha(r,t)} dt^{2} + e^{2 \beta(r,t)} dr^{2} + R(r,t)^{2} \,(d\theta^{2} + \sin^{2}\theta \,d\phi^{2})
\end{equation}
where $\alpha(r,t)$, $\beta(r,t)$ and $R(r,t)$ are the functions of the radial and time coordinates, and are to be determined from the solution of the field equations. The compact coordinates $\theta$ and $\phi$ are the standard angular coordinates of the round sphere cross- sections. In this frame of coordinates, the normalised velocity vector is given by $u^{\mu}=\exp[-\alpha(r,t)]\, (\partial/\partial t)^{\mu}$. It is useful to use this frame of coordinates since it provides easier integration of the Einstein equations: $G_{\mu\nu}=T_{\mu\nu}$ \footnote{In the following, we shall use the units $8\pi G=1$. One may view it as a rescaling of the fields appearing in the energy- momentum tensor and the cosmological constant, by appropriate factors. The cosmological constant shall be included in the energy- momentum tensor.} 
To incorporate the various forms of matter fields undergoing gravitational collapse in the universe, we envisage that the appropriate energy-momentum of the spherical ball of matter admits the following form:
\begin{equation}\label{Tuv}
T_{\mu\nu} = (p_t + \rho)u_{\mu} u_{\nu} + p_t g_{\mu\nu} + (p_r -p_t)x_{\mu} x_{\nu} - 2\eta \,\sigma_{\mu\nu} -\zeta \theta \,h_{\mu\nu} -\Lambda g_{\mu\nu},
\end{equation}
where $\rho$ is the energy density, $p_t$ and $p_r$ denote the tangential and radial pressure components, respectively. The quantities $\eta$ and $\zeta$ are the coefficients of shear and bulk viscosity, whereas $\sigma_{\mu\nu}$ is the shear tensor, $\theta$ is the expansion scalar, and $h_{\mu\nu}$ is the first fundamental form on the constant $t$ hypersurface and $x^{\mu}= e^{-\beta(r,t)} (\partial/\partial r)^{\mu}$ is a unit spacelike vector field tangential to the spacelike sections orthogonal to velocity vector $u^{\mu}$.
The expressions of the above quantities are 
\begin{eqnarray}
\theta &=& \nabla_{\mu} u^{\mu},~~~~~~h^{\mu}{}_{\nu}= (\delta^{\mu} _{\nu}+u^{\mu}\, u_{\nu}), \\
\sigma^{\mu\nu} &=&\frac{1}{2} (h^{\mu\lambda} \nabla_{\lambda} u^{\nu} + h^{\nu\lambda} \nabla_{\lambda} u^{\mu}) - \frac{1}{3} \theta h^{\mu\nu}.
\end{eqnarray}
Using the metric in the equation \eqref{guv}, these variables have the following values:
\begin{eqnarray}
\theta &=& e^{-\alpha} \left(\dot{\beta} + 2\dot{R}/{R} \right),~~~h_{\mu\nu} = e^{2 \beta(r,t)}dr^2 + R(r,t)^2(d\theta^2 + \sin^2\theta d\phi^2)\\
\sigma^{1}{}_{1}& =& (2/3)\left(\dot{\beta} - \dot{R}/R \right)e^{-\alpha},~~~
\sigma^{2}{}_2 = \sigma^{3}{}_3 = - (1/3)\left(\dot{\beta} - \dot{R}/R \right)e^{-\alpha}
\end{eqnarray}
It is standard to define the shear scalar $\sigma^{2} = (3/2)\sigma_{\mu\nu}\sigma^{\mu\nu} =e^{-\alpha} \left(\dot{\beta} - \dot{R}/{R} \right)$. 
In terms of these quantities, the non-zero components of the stress-energy-momentum tensor is now given by:
\begin{eqnarray}\label{T00}
T^{0}{}_{0} &=& -(\rho +\Lambda), ~~~~
T^{1}{}_{1} = p_r - \frac{4}{3}\eta \sigma - \theta \zeta - \Lambda,\\
T^{2}{}_{2} &=& p_t + \frac{2}{3}\eta \sigma - \theta \zeta - \Lambda,~~~~
T^{3}{}_{3} = p_t + \frac{2}{3}\eta \sigma - \theta \zeta - \Lambda.
\end{eqnarray}
In terms of the above mentioned fields, the independent components of the Einstein equations and the Bianchi identities useful for the study of gravitational collapse of matter fields are enumerated below:
\begin{eqnarray}\label{Einstein eq-1}
\rho &=& \frac{F'}{R^2 R^{\prime}},  \\
\label{Einstein eq-2}
 p_{r} &=& -\frac{\dot{F}}{R^2 \dot{R}} + \frac{4}{3}\eta \sigma + \theta \zeta,\\
\label{alpha-prime}
\alpha^{\prime}& = &\frac{2 R'}{R}\frac{p_t - p_r + 2\eta \sigma}{ \rho + p_r - (4/3)\eta \sigma - \zeta \theta} -  \frac{ \left[p_r -(4/3)\eta \sigma - \zeta \theta\right]^{\prime}}{ \rho + p_r - (4/3)\eta \sigma - \zeta \theta},  \\
\label{G01eqn}2\dot{R'} &=& R' \frac{\dot{G}}{G} + \dot{R} \frac{H'}{H},\\
\label{mass function}
F(r,t) & = &R\, [1 - G + H - (1/3)\Lambda R^{2}],
\end{eqnarray}
where  $G(r,t) = R^{'2} e^{-2\beta(r,t)}$, and $H(r,t) = \dot{R}^{2} e^{-2\alpha(r,t)}$, are two functions, and $F(r,t)$ is called the Misner- Sharp mass function. The equations \eqref{Einstein eq-1} and \eqref{Einstein eq-2}, are the $G_{00}$ and the $G_{11}$ equations respectively. The equation \eqref{alpha-prime} is the $r-$ component of the Bianchi identity $\nabla_{\mu}\, T^{\mu\nu}=0$. The equation \eqref{G01eqn}
is the $G_{01}$ component of the Einstein equation, while the last equation is the equation for mass function which gives the amount of mass contained in the spherical ball of matter of radius $R(r,t)$. As is well known, this is identical to the Kodama mass defined for a spherical ball of matter in the dynamically evolving spacetime. Before going to further discussions on the nature of these equations and regularity conditions on the initial data, it is useful to note that the change due to the introduction of the cosmological constant arises in the definition of the mass function $F(r,t)$, which in turn affects the equations for density and radial pressure, given in equations \eqref{Einstein eq-1} and \eqref{Einstein eq-2}. This is not surprising since the cosmological constant is known to provide a negative pressure, and contribute to the energy-momentum density. The other equations remain unaffected since they arise from the Binachi identities or from the off- diagonal term of the field equations. Interestingly, note that the very nature of these matter fields (including their contribution to pressure) and their appearance in the field equations opens up other possibilities, including that of a gravitational bounce at the late stage of gravitational collapse, as has been alluded to earlier. However, in the following, we shall only be interested in the continuous collapse of highly massive spherically symmetric self- gravitating matter configurations, for which, trapped surface formation and eventual formation of spacetime singularity is the only consequence depending on the initial data, correctness of the energy conditions and the nature of regularity conditions, as discussed below. \\

Note that several features are clear from these equations: First, from the equation \eqref{Einstein eq-1}, one observes that the density of matter diverges at $R(r,t)=0$, as well
as for $R^{\prime}(r,t)=0$. These two choices are clearly not identical: $R=0$ implies that the area radius of the matter configuration is vanishing, implying that the shells have all focussed to the centre, and therefore, this situation is also termed as creating a \emph{shell focussing} singularity. On the other hand, the choice of $R^{\prime}=0$ corresponds to the situation where shells of greater (or lesser) radius $R(r,t)$ cross shells of lesser (or greater) radius, and therefore, are referred to as \emph{shell crossing} singularities \cite{Yodzis:1973gha, Clarke, Christodoulou_1999, Dafermos:2008en}. Shell focussing singularities are real unavoidable singularities of spacetime, whereas shell crossing singularities are known to be gravitationally weak, and are generally coordinate artefacts. Indeed, it has been possible to construct spacetime extensions which removes such singularities. In the following, we shall not deal with shell crossing singularities, and all initial conditions have been chosen carefully so that such singularities donot arise during the evolution of the self- gravitating matter fields. Secondly, it is essential to determine the freely specifiable data for this physical set up. Clearly, the unknowns include the geometric variables in the metric $\alpha(r,t), \beta(r,t)$ and $R(r,t)$, and the matter variables $\rho(r,t), p_{t}(r,t), p_{r}(r,t)$, and the viscosity coefficients $\eta$ and $\zeta$. Given the five independent Einstein equations relating these variables, one is free to choose or specify three data. Several ways exist to provide such choices, and in the following, we shall show that given regular initial data, such data always leads to a well defined spacetime. The regularity conditions on the geometric and matter variables also follow from the equations above, and from any additional restrictions derived from the imposition of equations of state or from the energy conditions (along with the specification of the initial density and velocity distribution). For example, the usual way to ensure smoothness of initial data is to require that the metric functions be at least $C^{2}$. On the initial slice $t=t_{_{initial}}$, we require that $R(r,t_{_{initial}})=r$, and therefore, the eqn. \eqref{Einstein eq-1} leads to $\rho\simeq F^{\prime}\, r^{-2}$.
Hence, if the density has to be regular at the center of the matter configuration on the initial data slice, 
the mass function must admit a certain form given by $F(r,t)\simeq r^{3}\, m(t,r)$, where $m(t,r)$ is a regular function of $r$, and must be at least $C^{1}$. Similar conditions may be imposed on other functions as well, and further details may be found in \cite{joshi}.
\\

Before we dive into the main crux of the paper, it is important that
we connect the consequences of field equations to the behaviour of MTT. We have mentioned in the previous section that MTT does not have a specific signature, but carries the information as to how the horizons are developing in time. Let $\Delta$
denotes the MTT, a three dimensional surface in the manifold. Let $t^{\mu}=(\ell^{\mu}-C\, n^{\mu})$ be the
vector field which is orthogonal to the spherical cross-sections $\mathbb{S}^{2}$ 
and tangential to $\Delta$. Since $\ell\cdot n=-1$, the value of $C$ determines the signature of $t^{\mu}$. Since $t^{\mu}$ maps MTT to itself and it preserves the foliation, it follows that $\pounds_{t}\,\theta_{(\ell)}\triangleq 0$. Next, if
 $m^{\mu}$ and $\bar{m}^{\mu}$ are the null vector fields tangential to $\mathbb{S}^{2}$ (such that $m\cdot\bar{m}=1$), the area element is given by ${}^{2}\mathbf{\epsilon}=im\wedge\bar{m}$. The evolution of the area element of MTT is given by \cite{Dreyer:2002mx, Schnetter:2006yt}:
\begin{equation}\label{lietC}
\pounds_{t}\,{}^{2}\mathbf{\epsilon}=-C\,\theta_{(n)}\,{}^{2}\mathbf{\epsilon}
\end{equation}
It is easily seen from this equation that a timelike MTT (for which $C<0$) contracts, spacelike MTT (for which $C>0$) expands its area whereas a null MTT ($C=0$) does not grow in size. Also, it follows that
\begin{equation}
C=\frac{\pounds_{\ell}\,\theta_{(\ell)}}{\pounds_{n}\,\theta_{(\ell)}}.
\end{equation}
Using $\theta_{(\ell)}=0$ and the Einstein equation $G_{\mu\nu}=T_{\mu\nu}$, we get the following two equations:
\begin{eqnarray}
\pounds_{\ell}\,\theta_{(\ell)}&=&-T_{\mu\nu}\ell^{\mu}\ell^{\nu},\\
\pounds_{n}\,\theta_{(\ell)}&=&-(\mathcal{R}/2) +\Lambda+T_{\mu\nu}\ell^{\mu}n^{\nu}.
\end{eqnarray}
Here, $\mathcal{R}$ is the scalar curvature of the round $2$- sphere and may be rewritten
as $\mathcal{R}=(8\pi/\mathcal{A})$, where $\mathcal{A}$ is area of $2$- sphere.
These equations imply that the constant $C$ which determines the nature of the MTT is given by \cite{Booth:2006bn, Chatterjee:2020khj, Chatterjee:2024egb, Chatterjee:2021zre}:
\begin{equation}\label{value_of_c}
C=\frac{T_{\mu\nu}\ell^{\mu}\ell^{\nu}}{(4\pi/\mathcal{A})- \Lambda -T_{\mu\nu}\ell^{\mu}n^{\nu}}
\end{equation}
It follows from the discussion above that the signature of $\Delta$, determined by $C$,
is a quantity of utmost importance since it decides the nature and \emph{stability} of horizon
\cite{Andersson:2005gq, Andersson:2007fh}. 
From the above equation \eqref{value_of_c}, this value is controlled by 
the energy- momentum tensor and area of the marginally trapped
surfaces.\\


\section{Matter is Inhomogeneous and Pressureless} \label{section3}
In this section, we shall briefly discuss the process of continuous gravitational collapse of a dust configuration when the cosmological constant is positive. Note that the fluid is classified as dust if the coefficients of shear and bulk viscosity vanish $(\eta =\zeta=0)$ and
the tangential and radial components of pressure vanish, $p_{t} = p_{r} = 0$. Using eqn. \eqref{Einstein eq-1}, one gets
$ F' = \rho R^2 R'$, while the eqn. \eqref{Einstein eq-2} implies that $F = F(r)$. Further constraints arise from eqn. \eqref{alpha-prime}, which under the present assumptions, implies that $\alpha = \alpha(t)$. Defining a new coordinate time such that $ d{t} \rightarrow e^{\alpha}dt $, the function $\alpha$ may be absorbed. Further, the from these equations, it follows that $\dot{\beta} = \dot{R'}/{R'}$, and therefore we get $R^{\prime} = e^{\beta(r,t) + h(r)}$, where $h(r)$ is an arbitrary function. Redefining $ e^{2h(r)} = 1 - k(r)$, the equation for the mass function $F(r,t)$, in eqn. \eqref{mass function} reduces to:
\begin{equation}\label{eom}
\dot{R}(r,t)^2=\frac{F(r)}{R}-k(r)+\frac{1}{3} \Lambda R^2.
\end{equation}
In the following, we shall discuss the formation of the spacetime singularity, the EH, and discuss the formation of trapped surfaces as the matter shells collapse.
\subsection{Marginally Bound System}
The $k(r)=0$ signifies a collapse of marginally bound type for which, the velocity of the matter shells at the beginning of collapse is zero. The equation of motion \eqref{eom} for the collapsing shell reduces to
\begin{equation}\label{eom-mb}
\dot{R}(r,t)^2=\frac{F(r)}{R}+\frac{1}{3} \Lambda R^2.
\end{equation}
The solution of the above equation is given by
\begin{equation}\label{tRcurve_mb}
t = -\frac{2}{\sqrt{3 \Lambda}} \tanh^{-1}\left[\frac{(1/3) \Lambda R^3}{F(r)+ (1/3) \Lambda R^3} \right]^{1/2} + C,
\end{equation}
and describes the time curve $t(R)$ for the collapsing shell. The quantity $C$ is an arbitrary constant which may be fixed from the initial conditions: at $t=0$, the shells are at $R=r$, and therefore one gets:
\begin{equation}\label{constant}
C = \frac{2}{\sqrt{3 \Lambda}} \tanh^{-1}\left[\frac{(1/3) \Lambda r^3}{F(r)+ (1/3) \Lambda r^3}\right]^{1/2},
\end{equation}
This choice allows us to label the matter in the form of shells of fixed $r$, and track how the shells of a given $r$ evolves in time. For example, one may determine the time required for the shell to reach the singularity which can be determined by setting $R=0$ in eqn. \eqref{tRcurve_mb}. This implies that $t_{s}=C$, with its value as given in eqn. \eqref{constant}. Therefore, the shells of different $r$ reach the central singularity at different times. This is peculiar to inhomogeneous systems. On the other hand, if the configuration were homogeneous, and $F(r)=m_{0}\, r^{3}$, with $m_{0}$ a constant, a substitution in the equation \eqref{constant}, shows that the singularity time is independent of $r$, indicating therefore that it is same for all shells in the configuration. Getting back to the inhomogeneous configuration, if singularity time $t_{s}$ is shifted to $0$, the equation describing the motion of collapsing shell of radius $R(r,t)$ is given by:
\begin{equation}\label{R-shell}
R(t,r)= \left[\frac{3F}{\Lambda} \sinh^{2}{\biggl(\frac{\sqrt{3 \Lambda}t}{2}\bigg)} \right]^{1/3},
\end{equation}
It is better to rewrite the above equation in the standard format. The standard description of the matter collapse is given in terms of time curve $t(r,R)$, which is then obtained to be:
\begin{equation}\label{t-shell}
t(r,R)= \frac{2}{\sqrt{3 \Lambda}} \sinh^{-1}\left[\frac{\Lambda R^3}{3 F}\right]^{1/2}. 
\end{equation}
One may now use this equation to determine the time when the shell of labelled by a particular $r$ reaches its Schwarzschild- deSitter (or Kottler) radius $r_{H}$. The junction conditions in the appendix implies that
the mass function corresponding to this value of $r_{H}$
may be obtained through the relation: $F(r_H) = 2M=m(r_{H})\, r_{H}^{3}$, and is given by: 
\begin{equation}
r_{H}=\frac{2}{\sqrt{\Lambda}} \sin\left[\frac{1}{3} \sin^{-1}(3M \sqrt{\Lambda})\right]
\end{equation}
Therefore, the time $t_{H}$ for the shell labelled $r$ to reach its Kottler radius is obtained to be:
\begin{equation}\label{tH}
t_H = \frac{2}{\sqrt{3 \Lambda}} \sinh^{-1}\left[\frac{1}{M\sqrt{\Lambda}} \sin \,\left\{\frac{1}{3}  \sin^{-1}{(3M \sqrt{\Lambda})}\right\} - 1\right]^{1/2}.  
\end{equation}
Once the description of matter shells is fixed, whereby one ascertains the time for each shell to reach horizon, as well as the singularity, we now determine the location of the spherical trapped surfaces $R_{_{MTT}}$. These are obtained by solving the equations for the expansion of the null normals, given by $\theta_{(\ell)}=0$, and $\theta_{(n)}<0$, and in spherical symmetry, this is equivalent to the condition $ g^{ab} \nabla_{a} R\, \nabla_{b} R = 0$. This leads to a cubic equation in $R_{_{MTT}}$:
\begin{equation}\label{Eq-AH}
F(r,t) = R_{_{MTT}}-\frac{1}{3} \Lambda R_{_{MTT}}^3,
\end{equation}
Among the three solutions, two correspond to the locations of the black hole $(R_{BH})$ and the cosmological horizons $(R_{CH})$ respectively, while the third root is unphysical $(R_{U})$. Their values are as follows:
\begin{eqnarray}
    R_{CH}&=&\frac{Q_{A}}{\Lambda}+\frac{1}{Q_{A}}\\
    R_{BH}&=&\Bigl(-\frac{1}{2} - i \frac{\sqrt{3}}{2} \Bigr)\frac{Q_{A}}{\Lambda}+ \Bigl(-\frac{1}{2} + i \frac{\sqrt{3}}{2} \Bigr)\frac{1}{Q_{A}}\\
    R_{U}&=&\Bigl(-\frac{1}{2} + i\frac{\sqrt{3}}{2} \Bigr)\frac{Q_{A}}{\Lambda}+ \Bigl(-\frac{1}{2} - i \frac{\sqrt{3}}{2} \Bigr)\frac{1}{Q_{A}},
\end{eqnarray}
where $Q_{A} = \Lambda^{2/3} \left[-(3/2)F  + \sqrt{(9/4)\, F^2 -(1/\Lambda)}\right]^{1/3}$. The equation \eqref{R-shell} provides the time curve for spherically symmetric marginally trapped surface: 
\begin{equation}\label{R-ah}
 R_{_{MTT}}(r,t) = (3/ \Lambda)^{1/2}\,  \,\tanh\left[\frac{\sqrt{3 \Lambda}}{2}\, t\right].
\end{equation}
The equation describing the time development of EH is
also similarly obtained, from the time evolution of $R(r,t)$ along a radial null geodesic. The radial null geodesic of the outgoing photons implies that:
\begin{equation}
\biggl(\frac{dr}{dt}\biggr)_{Null} = \frac{1}{R'}
\end{equation}
Using $dR/dt = R^{\prime}\left(dr/dt\right)_{Null} + \dot{R}$, the time cuve for event horizon simplifies to
\begin{equation}
\frac{dR(t,r)}{dt} = 1 - \left[\frac{F(r)}{R}+\frac{\Lambda}{3}\, R^{2}\right]^{1/2}.
\end{equation}
Now, the equation \eqref{R-shell} implies that $F(r,t) = \left[(\Lambda R^{3}/3) \sinh^{-2}(\sqrt{3 \Lambda}/2)\, t \right]$.
As a result, the previous equation implies that the equation governing the time evolution of event horizon is given by:
\begin{equation}\label{t-EH eq}
\frac{dR}{dt} + \sqrt{\frac{ \Lambda }{3}} \coth\biggl(\frac{\sqrt{3 \Lambda}t}{2}\biggr)R = 1. 
\end{equation}
The solution may be computed using Mathematica. The version 13.3  gives the following 
solution:
\begin{equation}\label{t-EH sol-1}
R(t,r) = (2/5)\sqrt{3/\Lambda}\,~ \, \mbox{Hypergeometric}\, 2F1\bigl[1/2,5/6, 11/{6}, -\sinh^2{({\sqrt{3 \Lambda}/2}})t\bigr] \sin\left(\sqrt{3 \Lambda}/2  \right)t + C1,
\end{equation}
where $C1$ is a constant of integration, which may be fixed as follows: Since the event horizon is the last null ray reaching the null infinity, it must reach the horizon just at the same time as the shell $r$ reaches its Kottler radius, given by equation \eqref{tH}.
The expression for $C$ is obtained upon using equation \eqref{tH} in equation \eqref{t-EH sol-1}. Its value is not tidy and so, we refrain from showing it here, but shall use it for the plots below.\\

\textbf{Examples:} In the following, we consider several examples of gravitational collapse of matter with various initial density profile in the marginally bound models which lead to a black hole and a cosmological horizon. In each case, we have ensured, by choosing the parameters carefully, that the initial data has no (future) trapped surface to begin with. This implies the following: The condition for MTT is given by  $ g^{ab} \nabla_{a} R\, \nabla_{b} R = 0$, which leads to a cubic equation in $R_{_{MTT}}$, given in eqn. \eqref{Eq-AH}. On the initial data, the parameter choices ensure that the mass contained in any shell of radius $r$ must be smaller than $F(t,r)$, that is
$r < R_{BH}$. These choices of parameters also ensure that
no shell- crossing singularities may appear during the occurrence of gravitational collapse. \\
%

\begin{enumerate}
\item Let the initial mass density profile be of the form, with the parameters $\Lambda = 0.1,$ $m_0 = 1/(3\sqrt{\Lambda})$, and  $r_0 = 100\, m_0$.
\begin{equation}\label{k_0_theta}
\rho_0(r) = \frac{3 m_{0} }{500 \pi}\, \Theta(5 - r),
 \end{equation}
where $\Theta(x)$ is the Heaviside theta function. The graphs are plotted in fig. \eqref{k_0_theta_density_Rt}, and the nature of MTT is obtained in fig. \eqref{k_0_theta_c}.
\begin{figure}[tbh]
\begin{subfigure}{.45\textwidth}
\centering
 \includegraphics[width=6.5cm]{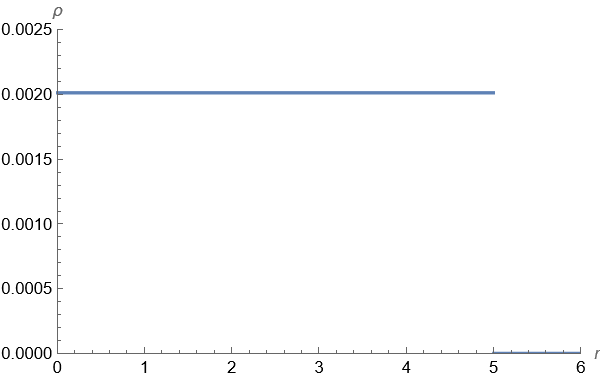} %
 \caption{}
\end{subfigure}
\qquad
\begin{subfigure}{.52\textwidth}
\centering
 \includegraphics[width=7.5cm]{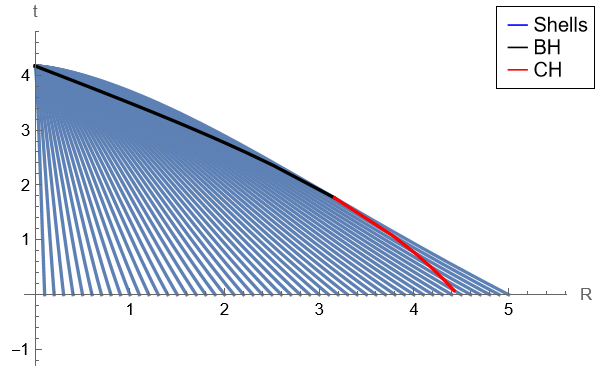}%
 \caption{}
\end{subfigure}
\caption{The figure corresponds to the initial density distribution referred in eq. \eqref{k_0_theta}. The graph (a) shows the density distribution, and corresponds to the OSD like model. The graph (b) combines many features: the blue lines show the collapse of matter shells marked by their radius on the initial slice $t_{initial}$. Therefore, the shells may be identified by their $r$ values. Note that since the density is a constant, all shells collapse to reach the central singularity at the same time, as has been argued previously in this section. The choice of parameters ensure that shells do not cross each other and hence no shell crossing singularities exist. The red curve in (b) is the  MTT corresponding to a cosmological horizon (CH), whereas the black curve is the MTT due to the black hole horizon (BH). Note that the MTT for BH is in accordance to the well known feature of the OSD model.  In the presence of the cosmological horizon, the black hole horizon seems to bifurcate from a common $R(r,t)$ of the MTT. }%
\label{k_0_theta_density_Rt}%
\end{figure}
%
\begin{figure}[tbh]
\begin{subfigure}{.52\textwidth}
\centering
 \includegraphics[width=6.5cm]{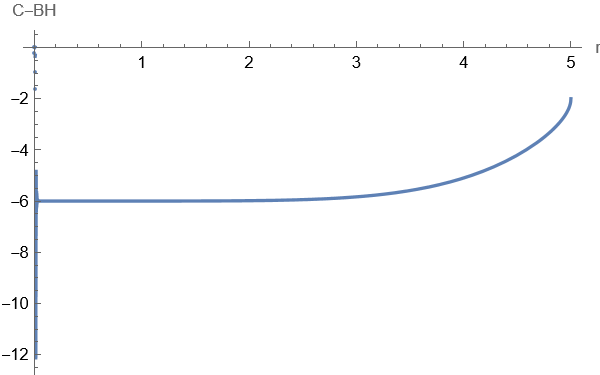}  %
 \caption{}
\end{subfigure}
\qquad
\begin{subfigure}{.48\textwidth}
\centering
 \includegraphics[width=6.5cm]{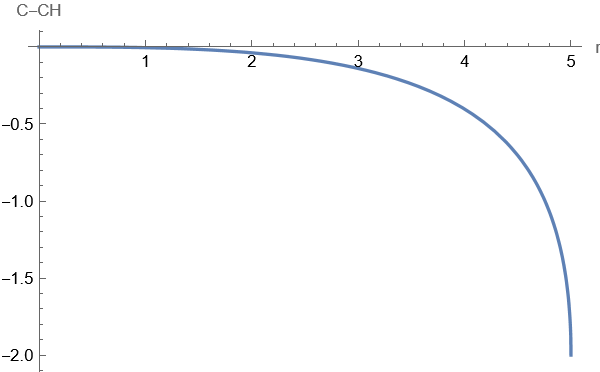}%
 \caption{}
\end{subfigure}
    \caption{ The graphs show the values of $C$ corresponding to the BH and the CH for the density distribution in eqn. \eqref{k_0_theta}. The negative values of $C$ implies, from 
    \eqref{value_of_c} and \eqref{lietC} that the MTTs are both timelike and hence, are unstable.}%
    \label{k_0_theta_c}%
\end{figure}

\item  Let us assume that the initial mass density is of the following form, with $\Lambda = 0.1$, $m_0 = 1/(3\sqrt{\Lambda})$, $r_0 = 100 m_0$. The graphs are plotted in figs. \eqref{k_0_Gaussian_Rt}, and \eqref{k_0_Gaussian_C}.
 \begin{equation}\label{density_gaussian}
 \rho_0(r) = \frac{m_0}{\pi^{3/2} r_0^{3}} e^{-r^2/r_0^{2}},
 \end{equation}
 %

%
\begin{figure}[tbh]
\begin{subfigure}{.52\textwidth}
\centering
 \includegraphics[width=5.5cm]{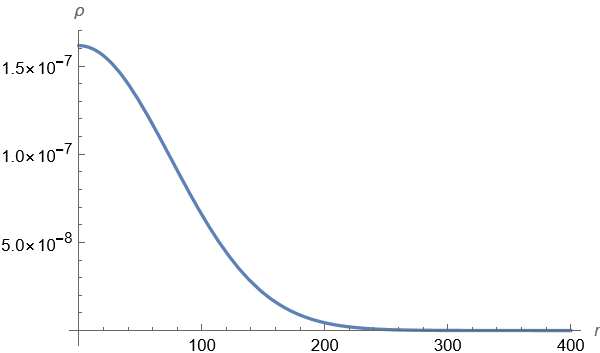} %
 \caption{}
\end{subfigure}
\qquad
\begin{subfigure}{.48\textwidth}
\centering
 \includegraphics[width=5.5cm]{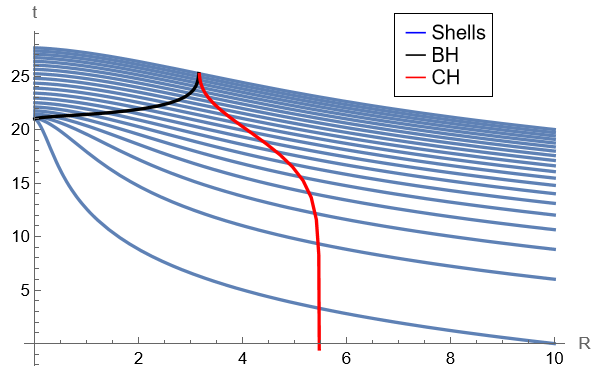}%
 \caption{}
\end{subfigure}
    \caption{ The figure in (a) shows the density profile of the matter density given in \eqref{density_gaussian}. The figure (b) shows the process of gravitational collapse by explicitly plotting the matter shells (in blue). Note that since the matter distribution is inhomogeneous, unlike the previous example, the matter shells do not reach the central singularity at the same time. The graph show the evolution of the MTTs corresponding to the black hole and the cosmological horizons. Note that while the black hole horizon grows in size, the radius of the cosmological horizon decreases in time, and matter shells cross. The MTTs eventually meet at the Nariai limit, given by $1/\sqrt{\Lambda}$. Note that the central singularity is hidden from the asymptotic observer. }%
 \label{k_0_Gaussian_Rt}%
    \end{figure}
\begin{figure}[tbh]
\begin{subfigure}{.52\textwidth}
\centering
 \includegraphics[width=6.5cm]{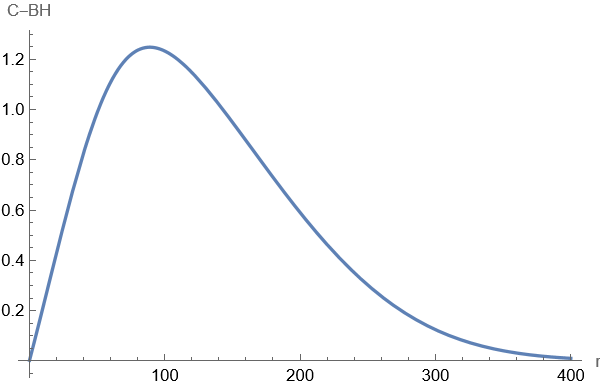} %
 \caption{}
\end{subfigure}
\qquad
\begin{subfigure}{.52\textwidth}
\centering
 \includegraphics[width=6.5cm]{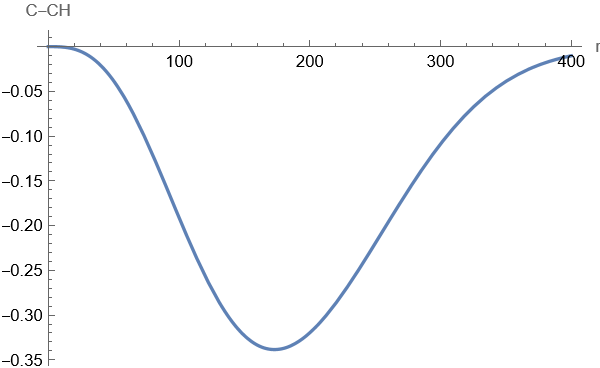}%
 \caption{}
\end{subfigure}
    \caption{The graphs show the nature of MTTs for the example in \eqref{density_gaussian}. Note that the MTT for black hole is spacelike whereas that of the cosmological is timelike. This is consistent with their behaviour depicted in fig. \eqref{k_0_Gaussian_Rt}(b). As the collapse process slows down or stops, the signature of the horizon becomes null. This implies that the horizon goes from being in the dynamical horizon phase to the isolated horizon phase.}%
      \label{k_0_Gaussian_C}%
\end{figure}
\item  The initial density of the following form, with parameters $\Lambda = 0.1$, $m_0 = 1/(3\sqrt{\Lambda})$, $r_0 = 10 m_0$. Again, the parameter choices ensure that no shell crossing singularities exist during the process of gravitational collapse. The plots are in fig. \eqref{k0_exp_Rt}, and fig. \eqref{k0_exp_C}.
\begin{equation}\label{k0_exp_density}
     \rho_{0}(r) = \frac{m_0}{8 \pi r_0^{3}} e^{-r/r_{0}}
\end{equation}
%

\begin{figure}[tbh]
\begin{subfigure}{.48\textwidth}
\centering
 \includegraphics[width=6.5cm]{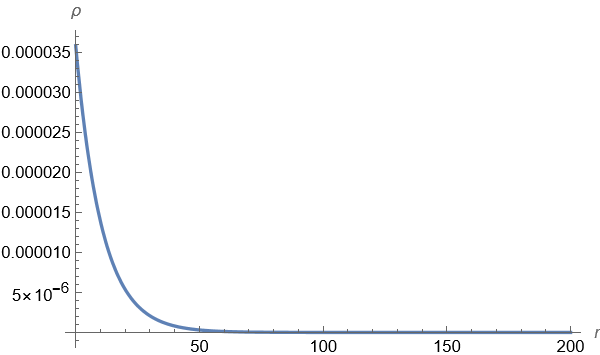} %
 \caption{}
\end{subfigure}
\qquad
\begin{subfigure}{.52\textwidth}
\centering
 \includegraphics[width=6.5cm]{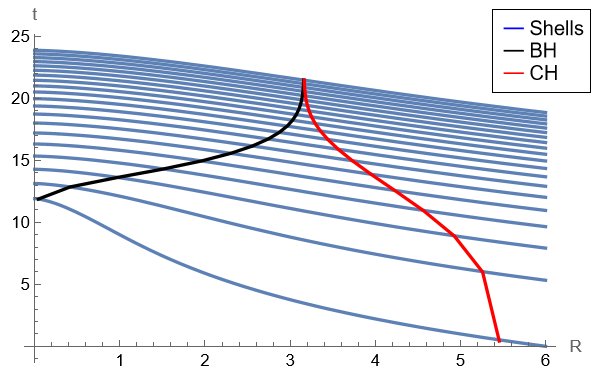}%
 \caption{}
\end{subfigure}
     \caption{The figure in (a) shows the density profile of the matter density given in \eqref{k0_exp_density}. The figure (b) shows the process of gravitational collapse by explicitly plotting the matter shells (in blue). Note that there is no shell- crossing singularity. The matter distribution is inhomogeneous
     and therefore the matter shells do not reach the central singularity at the same time. The evolution of the MTTs corresponding to the black hole and the cosmological horizons are also given. It shows that the black hole horizon grows in size, while the radius of the cosmological horizon decreases in time. The MTTs eventually meet at the Nariai limit, given by $1/\sqrt{\Lambda}$. }%
     \label{k0_exp_Rt}%
    \end{figure}
\begin{figure}[tbh]
\begin{subfigure}{.5\textwidth}
\centering
 \includegraphics[width=6.5cm]{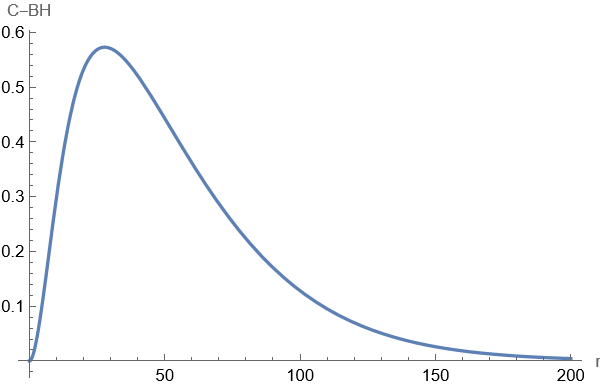} %
 \caption{}
\end{subfigure}
\qquad
\begin{subfigure}{.5\textwidth}
\centering
 \includegraphics[width=6.5cm]{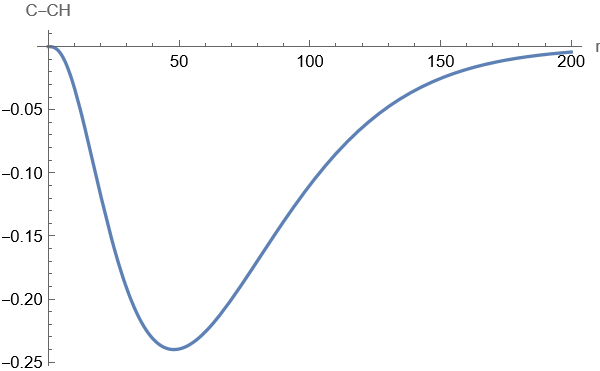}%
 \caption{}
\end{subfigure}
\caption{The nature of MTTs are plotted here. The black hole MTT is spacelike while that of the cosmological one is timelike. Again, this behaviour is consistent with those plotted in fig. \eqref{k0_exp_Rt}. As the collapse process slows down or stops, the signature of the horizon becomes null. This implies that the horizon goes from being in the dynamical horizon phase to the isolated horizon phase.}%
    \label{k0_exp_C}%
\end{figure}
\item In this example, we take the initial mass density of the following form, with $ \Lambda = 0.1,$ $m_0 = 1/(3\sqrt{\Lambda})$, $r_0 = 2 m_0$, $\sigma = 1$, 
 and $\varepsilon = 
 3 \sigma^3 \left[2 \pi \sigma (2 \sigma^2 + 3)(1 + \operatorname{erf}(\sigma)) + 4 \sqrt{\pi} e^{-\sigma^2} (1 + \sigma^2)
 \right]^{-1}$. With this choice of parameters, the graphs are given in fig. \eqref{k0_erf_Rt}, and fig. \eqref{k0_erf_C}.
%
 \begin{equation}\label{k0_erf_density}
 \rho_{0}(r) = \frac{m_0 \varepsilon}{r_0^3} \left[ 1 - \operatorname{erf} \left\{ \sigma \left( \frac{r}{r_0} - 1 \right) \right\} \right]  
\end{equation}
\begin{figure}[tbh]
\begin{subfigure}{.48\textwidth}
\centering
 \includegraphics[width=6.5cm]{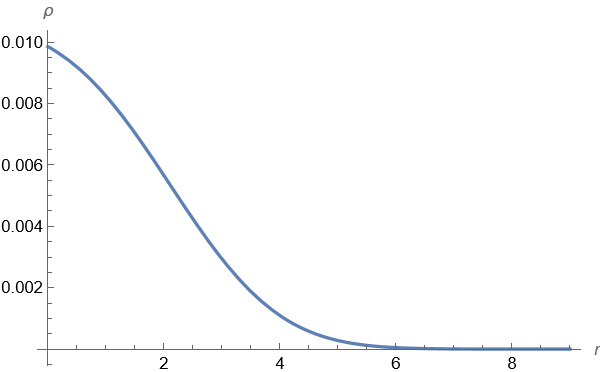} %
 \caption{}
\end{subfigure}
\qquad
\begin{subfigure}{.52\textwidth}
\centering
 \includegraphics[width=6.5cm]{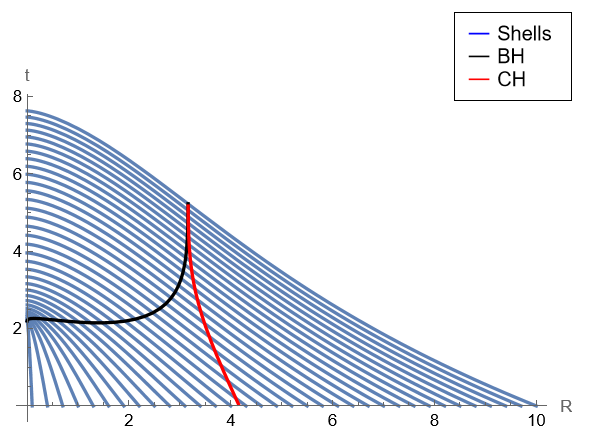}%
 \caption{}
\end{subfigure}
     \caption{The figure in (a) shows the density profile of the matter density given in \eqref{k0_exp_density}. For the choices of parameter values given above, the initial density profile is inhomogeneous. The figure (b) shows the matter shells (in blue). The matter shells do not reach the central singularity at the same time. The evolution of the MTTs corresponding to the black hole and the cosmological horizons show that the black hole horizon grows in size, while the radius of the cosmological horizon decreases in time. The MTTs eventually meet at the Nariai limit, given by $1/\sqrt{\Lambda}$.}%
     \label{k0_erf_Rt}%
 \end{figure}
\begin{figure}[tbh]
\begin{subfigure}{.52\textwidth}
\centering
 \includegraphics[width=6.5cm]{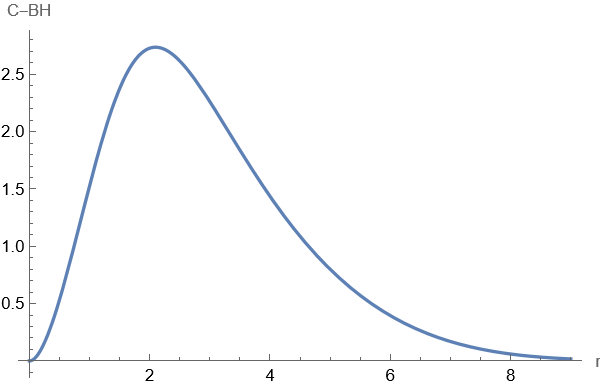} %
 \caption{}
\end{subfigure}
\qquad
\begin{subfigure}{.52\textwidth}
\centering
 \includegraphics[width=6.5cm]{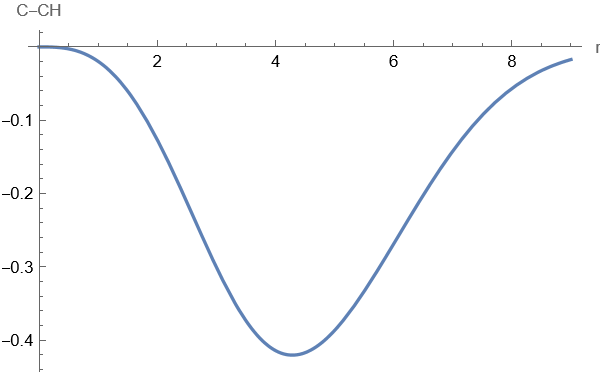} %
 \caption{}
\end{subfigure}
  \caption{The MTT corresponding to the black hole is spacelike while that of the cosmological is timelike. Again, this behaviour is consistent with those plotted in fig. \eqref{k0_erf_Rt}.}%
      \label{k0_erf_C}%
\end{figure}
\item The initial mass density of the following form with $\Lambda=0.01,$ $F = 2.22$, $m_0 = (F/2)$, $r_{j} = 2.5 F$, $r_0 = 10F$, $\alpha = 10.$ The figures are plotted in figs. \eqref{cobh_Rt_C}. This situation corresponds to the matter falling on an existing black hole. 
\begin{equation}\label{cobh_density}
\rho_{0}(r) = \frac{m_0}{2 \pi^{3/2} r_0^3 (2\alpha^2 + 1)} \cdot \exp\left[-\left(\frac{r}{r_0} - \alpha\right)^2\right]   \end{equation}
\begin{figure}[tbh]
\begin{subfigure}{.28\textwidth}
\centering
 \includegraphics[width=6.5cm]{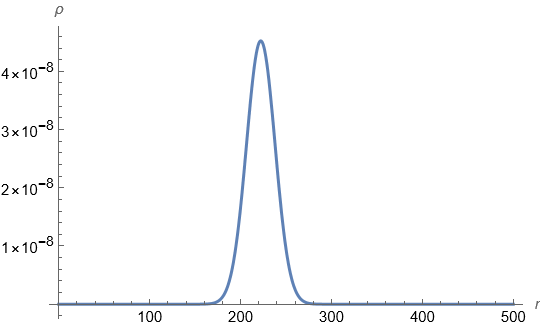} %
 \caption{}
\end{subfigure}
\qquad
\begin{subfigure}{.3\textwidth}
\centering
 \includegraphics[width=6.5cm]{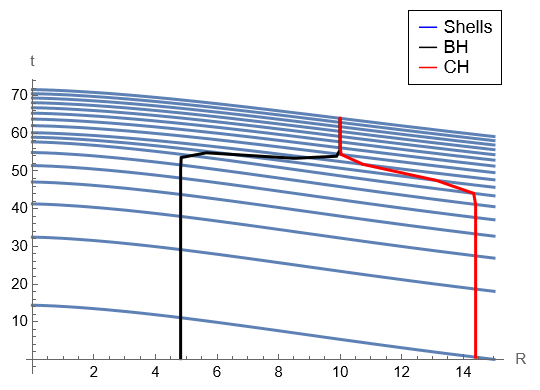} %
 \caption{}
\end{subfigure}
\qquad
\begin{subfigure}{.32\textwidth}
\centering
\includegraphics[width=5.4cm]
{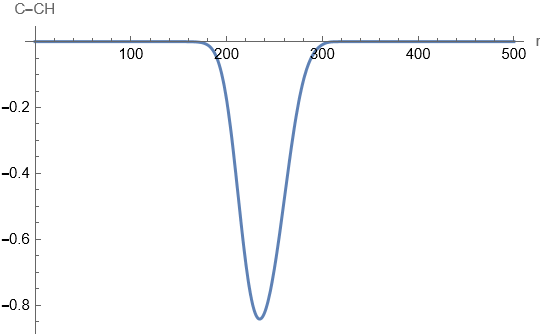} 
\end{subfigure}
    \caption{The plots corresponding to eqn. \eqref{cobh_density} is given here (a). Since a black hole already exists, the MTT for the black hole exists even before the matter has started to fall (black line), see fig (b). As more matter falls of appreciable density, the black hole starts to grow while the cosmological horizons begins to decrease so as to eventually meet at the Nariai limit $1/\sqrt{\Lambda}$. The figure (c) shows that the cosmological horizon is timelike. }
\label{cobh_Rt_C}
\end{figure}
%
\end{enumerate}

\subsection{Gravitationally bound configurations}
The $k(r)>0$ in eqn. \eqref{eom} signifies \emph{bound collapse} for which the velocity of matter shells at the beginning of dynamics is negative. The equation of motion of collapsing shell \eqref{eom} for $k(r) \neq 0$ is:
\begin{equation}\label{eom-b}
\dot{R}(r,t)^2 = -k(r) + \frac{F(r)}{R} + \frac{1}{3} \Lambda R^2.
\end{equation}
This equation may be solved analytically as discussed in \cite{Omer, Kraniotis:2001py, DAmbroise:2012cus}. The detailed solution is in the appendix \eqref{Appendix2}. The solution may be given in terms of parametric equations: 
\begin{equation}
R(\eta,r) = \frac{1}{[\,\mathcal{P}(\eta) - \delta(r)\,]}\,\left[\,\frac{\,F(r)}{4}\right]^{1/3},
\end{equation}
where the function $\delta(r)$ is given by:
\begin{equation}\label{deltar}
\delta(r)=(-k/3)\,[2 F(r)]^{-2/3}
\end{equation}
while the invariants of Weierstrass elliptic function $\mathcal{P}(\eta)$ are given by:
\begin{equation}
g_{2} = 12\, \delta(r)^{2}, ~~~~ g_{3} = -8\,\delta(r)^{3}-(\Lambda/3).
\end{equation}
The equation describing the relation between the time function $t$ and
the variable $\eta$ is given by (see the appendix \eqref{Appendix2} for details):
\begin{equation*}
t = \frac{1}{\mathcal{P'}(\epsilon)}\, \left[\ln \left\{\,\frac{ \sigma(\eta - \epsilon)}{\sigma(\eta + \epsilon)}\,\right\} + 2 \zeta(\epsilon) \,\eta \right],   
\end{equation*}
where $ \mathcal{P}(\epsilon)$ is taken equal to the function $\delta(r)$ in eqn. \eqref{deltar}, while $\zeta(\epsilon)$, and $\sigma(\eta)$ are the Weierstrass zeta and sigma functions \cite{Gradshteyn:1943cpj,Chandrasekharan,Whittaker_watson}. Here, the invariants $g_{2}$  and $g_{3}$ of the elliptic function $\mathcal{P}(\epsilon)$ remains undetermined, but may be chosen from the solution of the equation 
\begin{eqnarray}
\int \frac{1}{\sqrt{4z^{3} -g_{2}\, z-g_{3}}}\, dz =\epsilon. 
\end{eqnarray}
The solution of equation \eqref{eom-b} may also be derived using Mathematica, to obtain an expression for the $t- R$ curve, and is given as follows:
\begin{eqnarray}\label{math_sol-b}\label{bound_tR_curve_mathematica}
 t =\frac{2 \sqrt{3 R}}{\mathcal{A}} \left[ \mbox{EllipticF}\left\{\sin^{-1}\frac{R(-x+z)}{z(R-x)},\frac{(x-y)z}{(x-z)y} \right\}\,  - \mbox{EllipticPi}\left\{\frac{z}{z-x},   \sin^{-1} \frac{R(-x+z)}{z(R-x)}, \frac{(x-y)z}{(x-z)y} \right\} \, \mathcal{B}\right], 
\end{eqnarray}
where
\begin{equation}
  \mathcal{A}^{2}=  \left[3 F-3 k(r) R + \Lambda R^{3}\right] \, \left[\frac{xR(R-z)(x-z)}{(R-x)^2 }\right], ~~\mathcal{B}= \frac{x^{2}(R-z)(R-y)}{y(R-x)},
\end{equation}
with $\mbox{EllipticF}$ being the elliptic integral of the first kind, while $\mbox{EllipticPi}$ is the incomplete elliptic integral of the third kind. The quantities $x, y$ and $z$ are defined as follows
\begin{equation*}
 x = \frac{k}{\mathcal{X}} + \frac{\mathcal{X}}{\Lambda}, ~~y = -\frac{(1+ i \sqrt{3})k}{2\mathcal{X}} - \frac{(1- i \sqrt{3})\, \mathcal{X}}{2 \Lambda}, ~~z = -\frac{(1- i \sqrt{3})k}{2\mathcal{X}} - \frac{(1 +i \sqrt{3})\, \mathcal{X}}{2 \Lambda},
\end{equation*}
where $\mathcal{X}^{3}=[-(3/2)F\Lambda^{2} + \Lambda^{2}\sqrt{(9/4)F^{2}-(k^{3}/\Lambda)}]$.
This equation eqn. \eqref{bound_tR_curve_mathematica} may now be used to determine the $t-R$ curves for the collapse of shells. The examples are 
given below:\\

\begin{figure}[tbh]
\begin{subfigure}{.5\textwidth}
\centering
 \includegraphics[width=6.5cm]{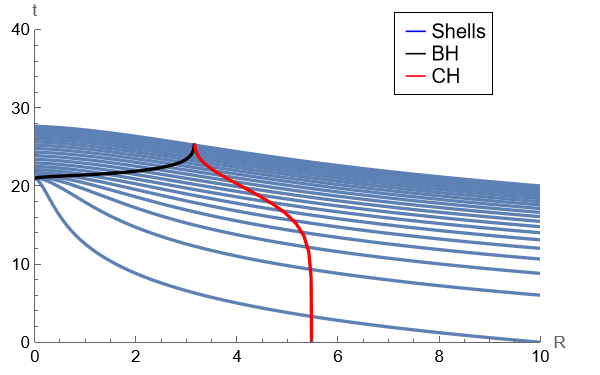} %
 \caption{}
\end{subfigure}
\qquad
\begin{subfigure}{.5\textwidth}
\centering
 \includegraphics[width=6.5cm]{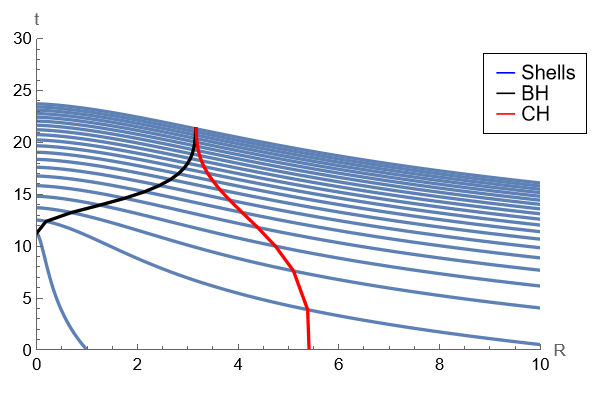} %
 \caption{}
\end{subfigure}
    \caption{The graph shows the formation of black hole and cosmological horizons along with shells, for the gaussian density distribution given in eqn. \eqref{density_gaussian} with parameters $\Lambda = 0.1,$ $m_0 = 1/(3\sqrt{\Lambda})$, $r_0 = 100 m_0$ and for exponential density distribution given in eqn. \eqref{k0_exp_density} with parameters $\Lambda = 0.1,$ $m_0 = 1/(3\sqrt{\Lambda})$, $r_0 = 10 m_0,$ respectively. As may be observed, the black hole horizon is spacelike while the cosmological horizon remains timelike in both cases. As the collapse process slows down or stops, the signature of the horizon becomes null. This implies that the horizon goes from being in the dynamical horizon phase to the isolated horizon phase.}%
     \label{Figure:14} %
\end{figure}

\textbf{Examples:}
Here, we consider two examples with the initial mass density given by the eqns. \eqref{density_gaussian} and \eqref{k0_exp_density} with values of $k$ appropriately chosen to ensure smooth and continuous curve. The parameters have been kept same to avoid shell crossings. The plots are in fig. \eqref{Figure:14}.


\section{Fluid matter with various properties including viscosity}\label{section4}
In the presence of the cosmological constant, the study of gravitational collapse of viscous matter becomes important in its own right. As has been explained in the introduction, most of matter in the universe may
have properties which mimic viscous damping. Therefore, the presence of such viscous fields naturally hinders the collapse process leading to an expected increase in the time for the shells to reach the central spacetime singularity, and may even delay the formation of trapped surfaces. In the following, we shall discuss these scenarios, and try to discern the behaviour of the shells accordingly.
%
\subsubsection{Fluids admitting tangential pressure only}
Let us begin with the description of gravitational matter for a fluid having only tangential pressure. In such a case, $p_{r}=0$ and $\eta=\zeta=0$. To close the system of equations, we assume an equation of state of the form $p_{t}=k_{t}\,\rho$. The equation for radial pressure 
eqn. \eqref{Einstein eq-2} for this system implies that:
\begin{equation}
\dot{F}(r,t) = 0, 
\end{equation}
which implies that the mass function is independent of time. 
From the equation eqn.\eqref{Einstein eq-1}, we have
\begin{equation}
\rho(r,t) = \frac{F(r)^{\prime}}{R^{2} R^{\prime}}  
\end{equation}
The $\alpha^{\prime}$ equation in eqn.\eqref{alpha-prime} also simplifies and now reads:
\begin{equation}\label{alpha_eqn_pt}
e^{2\alpha(r,t)} = R^{4\, k_{t}},
\end{equation}
whereas, the equation for the metric function $\beta(r,t)$ now takes 
the form
\begin{equation}\label{beta_eqn_pt}
e^{-2 \beta(t,r)} = \frac{k(r)R^{4 k_{t}}}{R^{\prime\, 2}}
\end{equation}
The equation for mass function $F(r)$ is obtained from \eqref{mass function}, and using the expressions for $\alpha(r,t)$ in eqn. \eqref{alpha_eqn_pt} and $\beta(t,r)$ in eqn. \eqref{beta_eqn_pt}is given by:
\begin{equation}
\dot{R}^2 = R^{4\, k_{t}} \left[\frac{F(r)}{R} - 1 + k(r)R^{4 \, k_{t}} + \frac{1}{3} \Lambda R^2 \right]    
\end{equation}
For simplification, we choose the parameter $k_{t} =-1/4$, and therefore
the 
\begin{equation}\label{tanp_shell_eqn}
R\, \dot{R} = - \left[F(r) - R + k(r) + \frac{1}{3} \Lambda R^{3}  \right]^{1/2}.   
\end{equation}
The parametric solution of the above equation is detailed in the appendix (Appendix-1), and is given by:
\begin{equation}\label{solution_R_eta_pt}
R(\eta,r)= \mathcal{P}(\eta + \epsilon)
\end{equation}
where $g_{2} = (12/\Lambda)$ and $g_{3} = -(12/\Lambda)\left[F(r)+k(r)\right]$ are invariants of Weierstrass elliptic function $\mathcal{P}(\eta + \epsilon)$ and $\epsilon$ 
is a constant. The variable $\eta$ is parametrically related to the time function $t$ through the following equation:
\begin{equation}
t = +\sqrt{(12/\Lambda)}~\, \zeta(\eta + \epsilon) + C ,
\end{equation}
where $\zeta(\eta)$ is the Weierstrass zeta function and C is the constant of integration, which may be fixed by requiring that at the initial time $t=0$, the constant $\epsilon=0$, and $R(r,0)=r$. In the following, we 
provide explicit examples for these scenarios by describing the collapse process with physically acceptable initial matter density profiles. \\

\begin{figure}[tbh]
\begin{subfigure}{.5\textwidth}
\centering
 \includegraphics[width=6.5cm]{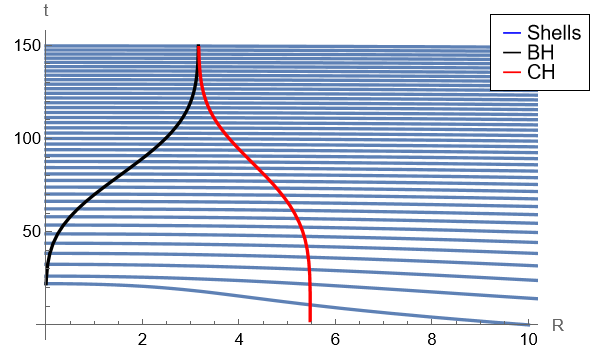} %
 \caption{}
\end{subfigure}
\qquad
\begin{subfigure}{.5\textwidth}
\centering
 \includegraphics[width=6.5cm]{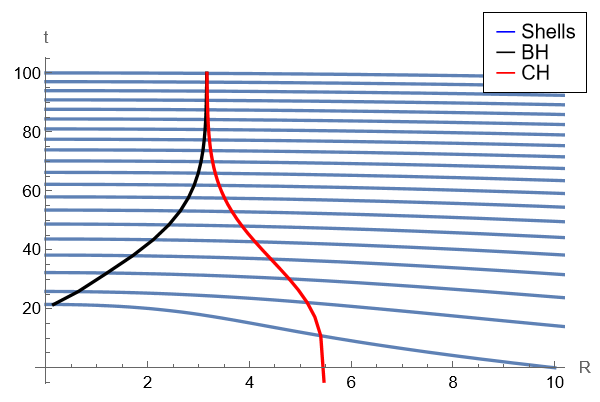} %
 \caption{}
\end{subfigure}
    \caption{The graph shows the formation of black hole and cosmological horizons along with shells, for the gaussian density distribution given in eqn. \eqref{density_gaussian} with parameters $\Lambda = 0.1,$ $m_0 = 1/(3\sqrt{\Lambda})$, $r_0 = 100 m_0$ and for exponential density distribution given in eqn. \eqref{k0_exp_density} with parameters $\Lambda = 0.1,$ $m_0 = 1/(3\sqrt{\Lambda})$, $r_0 = 10 m_0,$ respectively. In both cases, the MTTs match in the Nariai limit. }%
     \label{Figure:15}%
\end{figure}

\textbf{Examples:} To understand the process of gravitational collapse, we reconsider two simple scenarios from our previous section. The density distributions corresponding to the initial density distrbutions given in eqns. \eqref{density_gaussian} and
\eqref{k0_exp_density} are used to track the evolution of MTTs.
Again, the choice of parameters are dictated by the fact that no shell crossing singularities should occur and that no trapped surface should exist (corresponding to the black hole) on the initial Cauchy surface. The values of k are appropriately selected to ensure the smoothness of the resulting curve. The plots are in fig. \eqref{Figure:15}.

 %
 
\subsubsection{Fluids with radial pressure}
Let us now include the radial pressure into the viscous fluid, but to keep the discussion simpler and mathematically uncomplicated, we take the radial pressure of the form: $ p_{r} = \frac{4}{3}\eta \sigma + \theta \zeta$. We further assume, to close the system of equations, that the equation of state of the form $p_{t}= k_t \rho$, $p_{r} = k_r \rho$, $\theta= k_{\theta}\rho$ and $\sigma = k_{\sigma} \rho$. The solution proceeds along similar lines as the previous one: The equations \eqref{Einstein eq-1} and \eqref{Einstein eq-2} still gives: 
\begin{equation}
\rho = \frac{F'}{R^2 R'} , ~~~ \mbox{and} ~~~~ \dot{F}(t,r) = 0, 
\end{equation}
while the equations for $\alpha^{\prime}$ in \eqref{alpha-prime}, and $\dot{\beta}$
from the Bianchi gives respectively
\begin{equation}
e^{2\alpha(t,r)} = R^{4\, a_{1}}, ~~\mbox{and}~~~ e^{-2 \beta(t,r)} = \frac{k(r)R^{4a_{1}}}{R^{\prime\, 2}},
\end{equation}
where, $a_1= k_{t}+(2/3)\eta k_{\sigma}-\zeta k_{\theta}$.
From the mass function equation, \eqref{mass function}, we get by choosing the parameter $a_{1} =-1/4$, that the form of the equation becomes identical to those studied in the previous subsection:
\begin{equation}
R\, \dot{R} = - \left[F(r) - R + k(r) + \frac{1}{3} \Lambda R^{3}  \right]^{1/2}.   
\end{equation}
The parametric solution of the above equations is similar to those given in \eqref{solution_R_eta_pt}:
\begin{equation}
R(\eta,r)= \mathcal{P}(\eta + \epsilon)
\end{equation}
where $g_{2} = (12/\Lambda)$ and $g_{3} = -(12/\Lambda)\left[F(r)+k(r)\right]$ are invariants of Weierstrass elliptic function $\mathcal{P}(\eta + \epsilon)$ and $\epsilon$ 
is a constant. The variable $\eta$ is parametrically related to the time function $t$ through the following equation:
\begin{equation}
t = +\sqrt{(12/\Lambda)}~\, \zeta(\eta + \epsilon) + C ,
\end{equation}
where $\zeta(\eta)$ is the Weierstrass zeta function and C is the constant of integration, which may be fixed by requiring that at the initial time $t=0$, the constant $\epsilon=0$, and $R(r,0)=r$.\\ 

\begin{figure}[tbh]
\begin{subfigure}{.5\textwidth}
\centering
 \includegraphics[width=6.5cm]{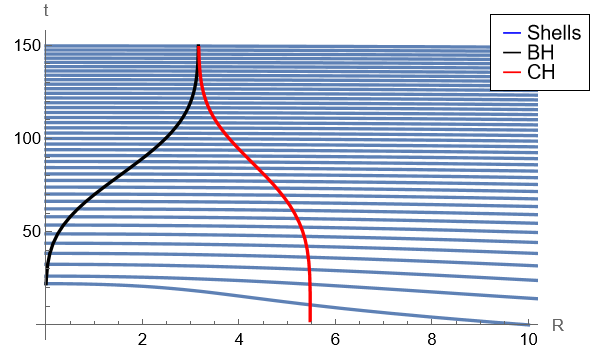} %
 \caption{}
\end{subfigure}
\qquad
\begin{subfigure}{.5\textwidth}
\centering
 \includegraphics[width=6.5cm]{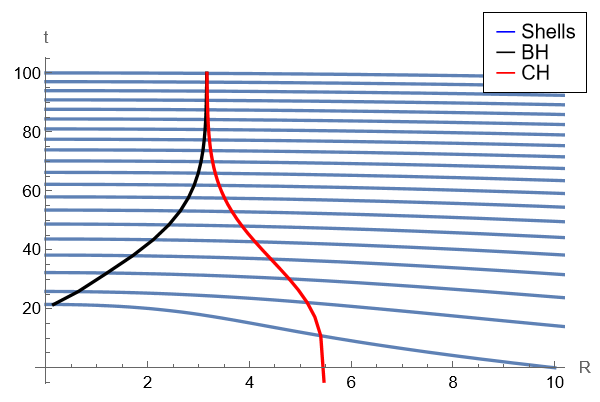} %
 \caption{}
\end{subfigure}
    \caption{The graph shows the formation of black hole and cosmological horizons along with shells, for the gaussian density distribution given in eqn. \eqref{density_gaussian} with parameters $\Lambda = 0.1,$ $m_0 = 1/(3\sqrt{\Lambda})$, $r_0 = 100 m_0$ and for exponential density distribution given in eqn. \eqref{k0_exp_density} with parameters $\Lambda = 0.1,$ $m_0 = 1/(3\sqrt{\Lambda})$, $r_0 = 10 m_0,$ respectively. The MTTs evolve to meet at the Nariai limit.  The central singularity is hidden from the asymptotic observer.}%
    \label{Figure_radial}%
\end{figure}

\textbf{Examples:} We consider the density distributions corresponding to the initial density distrbutions given in eqns. \eqref{density_gaussian} and
\eqref{k0_exp_density}. They are used to track the evolution of MTTs in fig. \eqref{Figure_radial}. The values of $k$ are appropriately selected to ensure smoothness of the resulting curve.
%
%
%

%
\subsubsection{Perfect fluids with equations of state}
A simplified solution is obtained for perfect fluids for which $p_t = p_r$ and $\eta=\zeta=0$. We further assume that the equations of state of the form $p_{t} = p_{r}=k_{r}\rho$ holds true. From the equations \eqref{Einstein eq-1}, and \eqref{Einstein eq-2}, we obtain the following simplified forms for density and radial pressure:
\begin{equation}
\rho(r,t) = \frac{F^{\prime}(t,r)}{R^2 R^{\prime}}, ~~\mbox{and},  ~~ p_{r} = -\frac{\dot{F}(t,r)}{R^{2} \dot{R}}.
\end{equation}
On the other hand, the expression for the metric coefficient in eqn. \eqref{alpha-prime}, with $a_{2} = [k_{r}/(1 + k_r)]$ implies that: $e^{2 \alpha(t,r)} =  \rho^{-2 a_2}$.
%
%
From the field equation eqn. \eqref{G01eqn}, we obtain:
$\left[\dot{G}/G\right]=(4\alpha^{\prime}\dot{R}/R^{\prime})$ wherefrom, we 
get that $\left[\partial\ln{G}/\partial t\right] =-4a_{2} (\rho^{\prime}/ \rho)\, (2\dot{R}/R^{\prime})$. This expression is easily integrated to:
\begin{equation}\label{Gint}
G(r,t) = 1 + r^{2}\, B(r,t), 
\end{equation}
where $B(r,t)$ is a smooth function. Next, from equation of mass function eqn. \eqref{mass function}, we get
\begin{equation}\label{fluid_eqnRdot}
\dot{R}^2 = \rho(r,t)^{-2 a_2} \left[\frac{F(r,t)}{R} + r^2 B(r,t) + \frac{1}{3} \Lambda R^2 \right]    
\end{equation}
Let us assume the following variables are of separable form: First is the mass function: $F(r,t) = F_{1}(r) F_{2}(t)$. Then, we also assume that the quantity $B(t,r)$ in eqn. \eqref{Gint} is 
$B(r,t) =  B_{1}(r) B_{2}(t)$. Also the shell radius $R(r,t) =  R_{1}(r) R_{2}(t)$, and the matter density may also be separated $\rho(r,t) =  \rho_{1}(r) \rho_{2}(t)$. Further, we choose very specific forms for these variables in relation to their time dependent parts:
$F_{2}(t) = \rho_{2}(t)^{2 a_2}$, $B_{2}(t) = -\rho_{2}(t)^{2 a_2}$ and
$R_{2}(t) = \rho_{2}(t)^{(2 a_2 -1)/3}$. Then the equation eqn. \eqref{fluid_eqnRdot} leads to
\begin{equation*}
\dot{R_{2}}(t) = -\frac{\rho_{1}(r)^{-a_2}}{R_{1}(r)^{3/2} \rho_{2}(t)^{(2 a_2 -1)/6}} \left[F_{1}(r) - r^2 B_{1}(r) R_{1}(r) \rho_{2}(t)^{(2 a_2 -1)/3} + (1/3) \Lambda R_{1}(r)^3 \rho_{2}(t)^{-1} \right]^{1/2}    
\end{equation*}
This equation may now be easily integrated to obtain the time curve $t=t(R)$. For further simplification, we choose the parameter $a_{2}=-1$, and obtain:
\begin{equation*}
dt = -\frac{R_{1}(r)^{3/2}}{\rho_{1}(r)} \frac{\sqrt{R_{2}(t)}}{\left[\, F_{1}(r) + \{- r^2 B_{1}(r) R_{1}(r) + \frac{1}{3} \Lambda R_{1}(r)^{3}\,\}\, R_{2}(t)\,\right]^{1/2}} dR     
\end{equation*}
The equation may be readily integrated with the boundary condition set to be: at $t=t_{i}=0$, $R(r,t)=r$ or $R(r,t)=R_{1}(r) R_{2}(t)=r$. The expression is of considerable length, but we quote it here for completeness.
\begin{equation}
t =(\mathcal{A}2/\mathcal{B}2)+\mathcal{C}2\, \tan^{-1}( \mathcal{D}2) + C
\end{equation}
where C is the constant of integration, and 
%
\begin{equation*}
\mathcal{A}2=[r^{3} R_{1} R_{2} (3 B_{1} r - \Lambda)]^{1/2}\,[9 F_{1} + 3 r^3 R_{1} R_{2} (-3 B_{1} r + \Lambda)]^{1/2} \,
\end{equation*}
\begin{equation}
   \times ([9 F_{1} + 3 r^3 R_{1} R_{2} (-3 B_{1} r + \Lambda)]^{1/2}+  6 F_{1} + r^3 R_{1} R_{2} [-3 B_{1} r + \Lambda)]) 
\end{equation}
%
\begin{equation*}
\mathcal{B}2=- r^3 \sqrt{R_{1}} (3 B_{1} r - \Lambda)^{3/2} \rho_{1}
\end{equation*}
\begin{equation}
     \times \left(-6 F_{1} + r^3 R_{1} R_{2} (3 B_{1} r - \Lambda) + 2 \sqrt{3 F_{1}} \sqrt{3 F_{1} + r^3 R_{1} R_{2} (-3 B_{1} r + \Lambda)}\right)
\end{equation}
\begin{equation}
\mathcal{C}2=6 F_{1} \left[6 \sqrt{3F_{1}} +\sqrt{3} r^3 R_{1} R_{2} (-3 B_{1}r + \Lambda) - 6\sqrt{F_{1}} \sqrt{3 F_{1} + r^3 R_{1} R_{2} (-3 B_{1} r + \Lambda)}\right]
\end{equation}
\begin{equation}
\mathcal{D}2={\frac{r^{3/2} \sqrt{R_{1}R_{2}} \sqrt{(3 B_{1} r - \Lambda_{1})}}{\left[-\sqrt{3F_{1}} + \sqrt{3 F_{1} + r^3 R_{1} R_{2} (-3 B_{1} r + \Lambda)}\right]}}
\end{equation}

\begin{figure}[tbh]
\begin{subfigure}{.5\textwidth}
\centering
 \includegraphics[width=6.5cm]{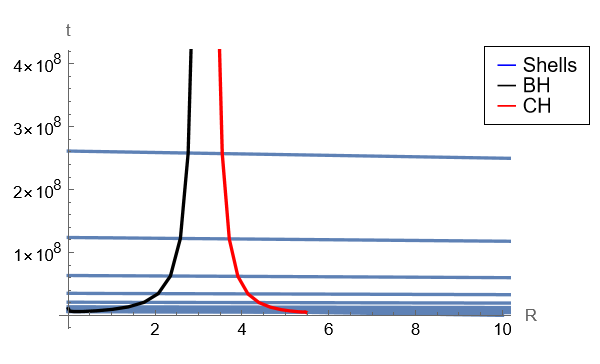} %
 \caption{}
\end{subfigure}
\qquad
\begin{subfigure}{.5\textwidth}
\centering
 \includegraphics[width=6.5cm]{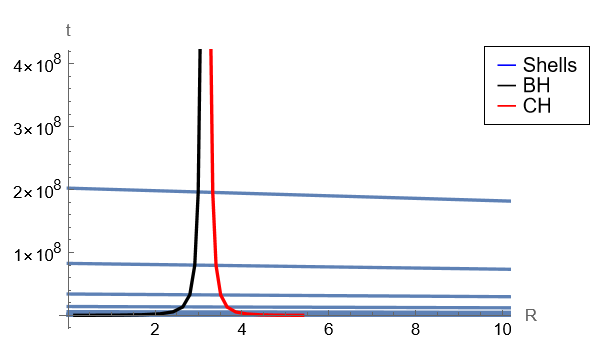} %
 \caption{}
 \end{subfigure}
    \caption{The graph shows the formation of black hole and cosmological horizons along with shells, for the gaussian density distribution given in \eqref{density_gaussian} with parameters $\Lambda = 0.1,$ $m_0 = 1/(3\sqrt{\Lambda})$, $r_0 = 100 m_0$ and for exponential density distribution given in \eqref{k0_exp_density} with parameters $\Lambda = 0.1,$ $m_0 = 1/(3\sqrt{\Lambda})$, $r_0 = 10 m_0,$ respectively. The Nariai limit is reached quite late and is not visible in the graph.}%
     \label{Figure:10}%
\end{figure}

\textbf{Examples:} In the following, we plot graphs for the collapse of matter shells, and locate the evolution of the black hole and cosmological horizons for the matter profiles in eqns. \eqref{density_gaussian} and \eqref{k0_exp_density}. The fluids
alter the time development of the MTTs, and one may note that although the parameters remain the same as before, the time required for the shells to reach singularity or the time for the MTTs to reach equilibrium is increased manifold. The graphs are in fig. \eqref{Figure:10}.
 %

\subsubsection{Viscous fluids with equations of state}
\begin{figure}[tbh]
\begin{subfigure}{.5\textwidth}
\centering
 \includegraphics[width=6.5cm]{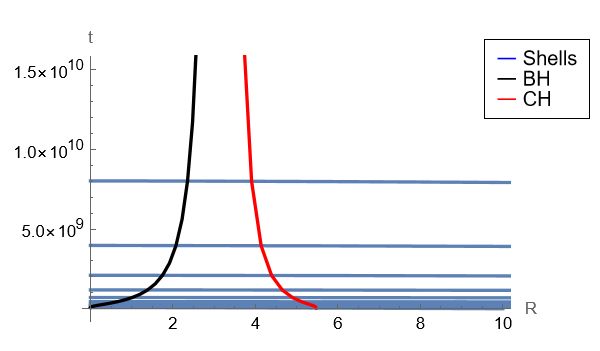} %
 \caption{}
\end{subfigure}
\qquad
\begin{subfigure}{.5\textwidth}
\centering
 \includegraphics[width=6.5cm]{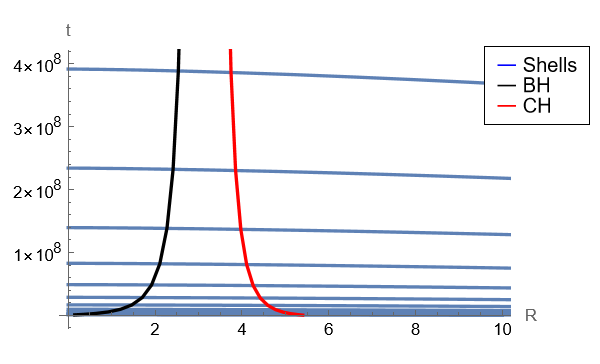} %
 \caption{}
 \end{subfigure}
   \caption{The graph shows the formation of black hole and cosmological horizons along with shells, for the gaussian density distribution given in \eqref{density_gaussian} with parameters $\Lambda = 0.1,$ $m_0 = 1/(3\sqrt{\Lambda})$, $r_0 = 100 m_0$ and for exponential density distribution given in \eqref{k0_exp_density} with parameters $\Lambda = 0.1,$ $m_0 = 1/(3\sqrt{\Lambda})$, $r_0 = 10 m_0,$ respectively.}%
      \label{Figure:11}%
\end{figure}

Using the field equations from eqn. \eqref{Einstein eq-1} and eqn. \eqref{Einstein eq-2}, the expressions for the density and radial pressure are given by:
\begin{equation}
\rho(r,t) = \frac{F(r,t)^{\prime}}{R^2 R'}, ~~~  
p_{r}(t,r) = -\frac{\dot{F}}{R^2 \dot{R}} + \frac{4}{3}\eta \sigma + \theta \zeta .
\end{equation}
The equation for the metric variable in equation \eqref{alpha-prime} gives the following:
\begin{equation}
\alpha' = \frac{2 R'}{R}\frac{p_t - p_r + 2\eta \sigma}{ \rho + p_r - (4/3)\eta \sigma - \zeta \theta} -  \frac{ (p_r -(4/3)\eta \sigma - \zeta \theta)'}{ \rho + p_r - (4/3)\eta \sigma - \zeta \theta}  
\end{equation}
Similar to the previous subsection, we assume equations of state: The tangential and radial pressures are separately related to the density, $p_t= k_t \rho$, and $p_r = k_r \rho$. Similarly, for simplicity, the geometric variables also are related: $\theta= k_{\theta}\rho$ and $\sigma = k_{\sigma} \rho$. If we define
$a_{1}= [k_t - k_r + 2\eta k_{\sigma}] [1 + k_r - (4/3)\eta k_{\sigma} - \zeta k_{\theta}]^{-1}$, and  $a_{2}= [(k_r -(4/3)\eta k_{\sigma} - \zeta k_{\theta})^{\prime}\,][ 1 + k_{r} - (4/3)\eta k_{\sigma} - \zeta k_{\theta}\, ] $, the equation eqn. \eqref{alpha-prime} leads to:
\begin{equation*}
e^{2 \alpha (t,r)} = R^{4 a_1}\,\rho^{-2 a_2}.
\end{equation*}
On the other hand, the eqn. \eqref{G01eqn} leads to:
\begin{equation*}
\frac{\partial}{\partial t}(\ln G)=2 \left[2 a_1 \frac{R'}{R} - a_{2} \frac{\rho'}{\rho} \right] \frac{2\dot{R}}{R'},  
\end{equation*}
which is easily integrated to give $G(t,r) = 1 + r^2 B(r,t)$. 
To get the equations governing the motion of matter shells, 
we use the equation for mass function \eqref{mass function}, to get: 
\begin{equation*}
\dot{R}^2 = R(r,t)^{4 a_1} \rho(r,t)^{-2 a_2} \left[\frac{F(r,t)+ r^2 B(r,t) R(r,t) + \frac{1}{3} \Lambda R(r,t)^3}{R(r,t)} \right]  
\end{equation*}
Again, we assume the functions to be of separable form just as before, and this gives:
\begin{equation*}
\dot{R_{2}}(t) = -\frac{R_{1}(r)^{2 a_1} R_{2}(t)^{2 a_1} \rho_{1}(r)^{-a_2}}{R_{1}(r)^{3/2} \rho_{2}(t)^{(2 a_2 -1)/6}} \left[F_{1}(r) - r^2 B_{1}(r) R_{1}(r) \rho_{2}(t)^{(2 a_2 -1)/3} + \frac{1}{3} \Lambda R_{1}(r)^3 \rho_{2}(t)^{-1} \right]^{1/2}    
\end{equation*}
To integrate, we choose, for simplification, the parameter $a_{2}=-1$ and $a_{1}=-1/4$, and obtain the time curve for the shells as
\begin{equation*}
t = \frac{2 R_{1}(r)^{2}}{\rho_{1}(r)} \frac{N\, \sqrt{M}}{Q} + C
\end{equation*}
where C is the constant of integration fixed through the boundary condition, $t=t_{i}=0$, $R(r,t)=r$ or $R(r,t)=R_{1}(r) R_{2}(t)=r$, and
\begin{equation}
\mathcal{N}=2 F_{1}(r) + r^2 B_{1}(r)R_{1}(r) R_{2}(t) + 1/3 R_{1}(r)^{3} R_{2}(t)
\end{equation}
\begin{equation}
\mathcal{M}=F_{1}(r)-R_{1}(r)R_{2}(t)\,[r^2 B_{1}(r)+1/3 R_{1}(r)^{2}], ~~~~~\mathcal{Q}=r^{2} B_{1}(r) R_{1}(r) + (1/3) R_{1}(r)^{3}.
\end{equation}
\textbf{Examples:}
In the following, we plot graphs for the collapse of matter shells, and locate the evolution of the black hole and cosmological horizons for the matter profiles in eqns. \eqref{density_gaussian} and \eqref{k0_exp_density}. Due to viscous effects in the fluid, the time development of the MTTs
slows down considerably and one may note that although the parameters remain the same as before, the time required for the shells to reach singularity or the time for the MTTs to reach equilibrium is increased by quite some few orders of magnitude, see fig. \eqref{Figure:11}.


\section{Discussions}
The main objective of this paper was to study the formalism of gravitational collapse of spherically symmetric matter distribution using the
Einstein field equations with a positive cosmological constant.
The central idea behind the calculation may be summarised as follows: 
First, the matter distribution is assumed to be in the form of 
spherical shells labelled by the value of its coordinate $r$ on 
the initial time slice $t=0$, implying that
on the initial slice, the value of the areal radius is $R(0,r)=r$. Then, using the Einstein field equations, time evolution of 
the areal coordinate $R(t,r)$ is determined. This is given, for example,
through the equation \eqref{eom} in situations where the matter
is distributed inhomogeneously. This equation gives the time evolution of collapsing shells, and in the text, the equations of this kind have been sometimes referred 
to as the shell equation. The solution of these equations
has been carried out under various choices of the mass profile, and shows how the radius of the collapsing shell decreases with time so as to reach the central spacetime singularity. In this paper, we have determined the time taken for the shells to reach the central singularity for a variety of density profiles. Further, in each case, we have used the formalism of marginally trapped tube to determine the formation of the black hole horizon and the evolution of both the black hole and cosmological horizon with time. In each of the situations studied in the paper, as matter falls through, the black hole horizon grows with time (with its signature being spacelike), whereas the radius cosmological horizon decreases with time (its signature being timelike during the process). These two horizons eventually meet in each case, at the so called Nariai limit. Our examples in the previous subsections provide explicit construction of these horizons 
and provides graphical demonstration of the process in some detail. Moreover, our construction provides, though the constant $C$, see equation \eqref{value_of_c}, the signature of the MTT and its stability.  
The second issue is related to the effect of matter properties and the cosmological constant on the collapse process itself. It may be seen quite readily from the graphs that the presence of tangential pressure and viscosity in the fluids alter the time of singularity formation, as well as increases the horizon formation time, as compared to dust collapse models. The presence of the positive cosmological constant introduces additional horizon and its development is equally affected by the presence of viscosity or fluid properties. In particular, the presence of viscosity alters the time taken for the shells to reach the central singularity by a significant factor.
Even the horizon formation time is altered dramatically. So, in a viscous de Sitter universe, the formation of compact objects need to be studied more carefully. The third issue is related to the formulation of MTT in understanding gravitational collapse. It is clear from our study that MTTs can indeed play the role of a horizon in this process. It exactly mimics the dynamics of horizon: as matter falls, the black hole horizon grows (and cosmological horizon decreases), and the horizons become null as the matter stops to fall. In all these cases, the spacetime singularity remains hidden from the asymptotic observers, and hence supports the censorship conjecture.\\

However, there are still some gaps in our present calculation, and our results, although exact in the analytical sense, depend on assumptions on the forms of equations of state and the values of the equations of state parameters. The values of these parameters usually happen to be those for which a solution is known and are expressed in a closed form, for example in terms of elementary or higher transcendental functions. In situations where values associated with these parameters are such that the exact solutions are not known, one has to resort to numerical solutions. Among other things, the difficulties associated with such numerical solutions comprise of lack of control. For example, in most of these cases, it becomes difficult to invert the equation $t=t(R)$ to obtain the relation $R=R(t)$. Therefore, the lack of exact solutions is an hindrance in obtaining an exact behaviour of matter shells under gravitational collapse. In this paper, we have obtained several exact solutions which may be used to study a large class of solutions, and may even be extended to solutions which are small perturbations thereof. \\

To summarise, the formalism of spherically symmetric gravitational collapse of matter fields in the deSitter spacetimes have been studied in some generality and studies other types of matter may be carried out as well, for the situations where the trapped surfaces are not spherical. That would lead to a complete study of the evolution of the black hole region and its boundary \cite{ Eardley:1997hk, BenDov:2006vw}.  


\section*{Acknowledgements}
The authors thank Prof. Amit Ghosh and Dr. Suresh Jaryal for discussions. AC is supported through the DAE-BRNS project $58/14/25/2019$-BRNS. 


\section{Appendix}
In the following appendixes, we collect several formula which have been 
used throughout the text. In the first appendix, we derive the field equations and 
the mass formula. The second and the third appendix, \ref{Appendix1} and \ref{Appendix2} solves the equations of motion analytically, through direct integration, and is expressed in terms of higher transcendental functions including the Weierstrass functions.

\subsection{Derivation of the field equations}

Let us consider a general spherically symmetric metric expressed in the form
\begin{equation}
ds^2 = -e^{2 \alpha(r,t)} dt^{2} + e^{2 \beta(r,t)} dr^{2} + R(r,t)^{2} (d\theta^{2} + \sin^{2}\theta \,d\phi^{2})
\end{equation}
where $\alpha$, $\beta$ and R are the functions of radial and time coordinates. The components of the Einstein tensor are given by
\begin{equation}\label{G00}
G_{00} = \frac{e^{2\alpha}}{R^2}+ \frac{2 \dot{R} \dot{\beta}}{R} + \frac{\dot{R}^2}{R^2} + e^{2\alpha} e^{-2\beta} \biggl(\frac{2 R' \beta'}{R} - \frac{2 R''}{R} - \frac{R^{'2}
}{R^2}\biggr),
\end{equation}
\begin{equation}\label{G01}
G_{01} = G_{10} = -\dot{R'} + \alpha' \dot{R} + \dot{\beta} R',
\end{equation}
\begin{equation}\label{G11}
G_{11} = -\frac{e^{2\beta}}{R^2}+ \frac{ \alpha' R'}{R} + \frac{R^{'2}}{R^2} - e^{-2\alpha} e^{2\beta} \biggl(-\frac{2 \dot{R} \dot{\alpha}}{R} + \frac{2 \ddot{R}}{R^2} + \frac{\dot{R}^2}{R^2}\biggr),
\end{equation}
\begin{equation}\label{G22}
G_{22} = e^{-2\beta} R(R''+R' \alpha'' - R'\beta' - R \alpha^{'2} -R \alpha' \beta') + e^{-2\alpha} R(-\ddot{R} + \dot{R} \dot{\alpha} - \dot{R} \dot{\beta} + R \dot{\alpha} \ddot{\beta} + R \dot{\beta} \ddot{\beta} - R \dot{\beta}^2 + R \dot{\alpha} \dot{\beta}),
\end{equation}
\begin{equation}\label{G33}
G_{33} = R \sin^2{\theta}[e^{-2\beta} (R''+R' \alpha'' - R'\beta' - R \alpha^{'2} -R \alpha' \beta') + e^{-2\alpha} (-\ddot{R} + \dot{R} \dot{\alpha} - \dot{R} \dot{\beta} + R \dot{\alpha} \ddot{\beta} + R \dot{\beta} \ddot{\beta} - R \dot{\beta}^2 + R \dot{\alpha} \dot{\beta})],
\end{equation}
The general form of the stress-energy-momentum tensor for the spherical ball of a viscous fluid is as follows:
\begin{equation}
T_{ab} = (p_t + \rho)u_a u_b + p_t g_{ab} + (p_r -p_t)X_a X_b - 2\eta \sigma_{ab} -\zeta \theta h_{ab} -\Lambda g_{ab},
\end{equation}
where $\rho$ is the energy density, $p_t$ and $p_r$ denote the  tangential and radial pressure components, respectively. The shear and bulk viscosity coefficients are given by $\eta$ and $\zeta$. $\sigma_{ab}$ is the shear tensor, $\theta$ is the expansion scalar, $h_{ab}$ is the projection tensor, and $X^a$ is a unit space-like vector tangential to the spacelike section orthogonal to velocity vector $u^a$, satisfying $X^a X_a = 1$. The expressions of the above quantities are 
\begin{equation}
\theta = \nabla_a u^a, ~~~
h^a_{b}= (\delta^a _b+u^a u_b),
\end{equation}
\begin{equation}
\sigma^{ab} = \frac{1}{2} (h^{ac} \nabla_c u^b + (h^{bc} \nabla_c u^a) - \frac{1}{3} \theta P^{ab},
\end{equation}
\begin{equation}
X^{a} = e^{-\beta(r,t)} \biggl(\frac{\partial}{\partial r} \biggr)^a, ~~~~
u^{a} = e^{-\alpha(r,t)} \biggl(\frac{\partial}{\partial t} \biggr)^a
\end{equation}
Their values determined from the metric are given below:
\begin{equation}
\theta = e^{-\alpha} \biggl(\dot{\beta} + 2\frac{\dot{R}}{R} \biggr),~~~~~~
h_{ab} = e^{2 \beta(r,t)}dr^2 + R(r,t)^2(d\theta^2 + \sin^2\theta \, d\phi^2),
\end{equation}
\begin{equation}
\sigma^1 _1 = \frac{2}{3}\biggl(\dot{\beta} - \frac{\dot{R}}{R} \biggr)e^{-\alpha},
~~~~
\sigma^2 _2 = \sigma^3 _3 = - \frac{1}{3}\biggl(\dot{\beta} - \frac{\dot{R}}{R} \biggr)e^{-\alpha},
\end{equation}
Defining shear scalar $ \Bar{\sigma} = \sigma_{ab} \sigma^{ab} $
gives:
$\Bar{\sigma}^2 = \frac{2}{3}\biggl(\dot{\beta} - \frac{\dot{R}}{R} \biggr)^2 e^{-2\alpha}$.
For simplification, we redefine $\sigma$ as $\sigma^2 = e^{-\alpha} \left(\dot{\beta} - \frac{\dot{R}}{R} \right)$. The non-zero components of the stress-energy-momentum tensor:
\begin{equation}\label{T0}
T_{00} = -e^{-2\alpha}(-\rho -\Lambda),~~~~
T_{11} = e^{-2\beta}(p_r - \frac{4}{3}\eta \sigma - \theta \zeta - \Lambda),
\end{equation}
\begin{equation}\label{T22}
T_{22} = R^2(p_t + \frac{2}{3}\eta \sigma - \theta \zeta - \Lambda),
~~~~~
T_{33} = R^2 \sin^2{\theta}\,(p_t + \frac{2}{3}\eta \sigma - \theta \zeta - \Lambda).
\end{equation}
From the Bianchi identities: $ \nabla_a T^{ab} = 0 $, we have the t-equation and the r-equation. The t-equation is given by 
\begin{equation}
\dot{\rho} e^{-\alpha} + (\rho + p_t)\theta + (p_r - p_t)\dot{\beta} e^{-\alpha}  - \frac{4}{3}\eta \sigma^2 - \theta \zeta^2 = 0,
\end{equation}
on rearranging
\begin{equation}
\dot{\beta} = -\frac{\dot{\rho}}{ \rho + p_r - (4/3)\eta \sigma - \zeta \theta} - \frac{2\dot{R}}{R} \frac{ \rho + p_t + (2/3)\eta \sigma - \zeta \theta}{ \rho + p_r - (4/3)\eta \sigma - \zeta \theta},
\end{equation}
and the r-equation is given by 
\begin{equation}
(p_t - \zeta \theta)' + \alpha'(\rho + p_t + \zeta \theta) - \frac{4}{3}\eta \sigma' - \frac{4}{3}\eta \sigma(\alpha' + 3R'/R) + (p_r - p_t)' + (\alpha' + 2R'/R)(p_r - p_t) = 0,
\end{equation}
on rearranging 
\begin{equation}
\alpha' = \frac{2 R'}{R}\frac{p_t - p_r + 2\eta \sigma}{ \rho + p_r - (4/3)\eta \sigma - \zeta \theta} -  \frac{ (p_r -(4/3)\eta \sigma - \zeta \theta)'}{ \rho + p_r - (4/3)\eta \sigma - \zeta \theta}.  
\end{equation}
Let us consider $G_{01}$ component of Einstein tensor
\begin{equation}
-\dot{R'} + \alpha' \dot{R} + \dot{\beta} R' = 0,
\end{equation}
Using $ \alpha'$ and $ \dot{\beta}$ in the above equation and multiplying by $R^2$, we get
\begin{equation}
[(p_r - (4/3)\eta \sigma - \zeta \theta)R^2 \dot{R}]_{,r} + [\rho R^2 R']_{,t} = 0.
\end{equation}
To interpret the exact differential, let us construct a function $F(r,t)$ such that
\begin{equation}
F' \propto \rho R^2 R' ,
\end{equation}
\begin{equation}
\dot{F} \propto -(p_r - (4/3)\eta \sigma - \zeta \theta)R^2 \dot{R},
\end{equation}
Determining the exact differential:
Equation $G_{00}$ given by,
\begin{equation}
\frac{e^{2\alpha}}{R^2}+ \frac{2 \dot{R} \dot{\beta}}{R} + \frac{\dot{R}^2}{R^2} + e^{2\alpha} e^{-2\beta} \biggl(\frac{2 R' \beta'}{R} - \frac{2 R''}{R} - \frac{R^{'2}
}{R^2}\biggr) =  -e^{-2\alpha}(-\rho -\Lambda),
\end{equation}
Multiplying $G_{00} $ by $R^2 R'$ the above equation reduces to
\begin{equation}
[R(1+ \dot{R}^2 e^{-2\alpha} + R^{'2} e^{-2\beta} - (1/3)\Lambda R^2 )]_{,r} = \rho R^2 R'.
\end{equation}
Equation $ G_{11} $ is given by
\begin{equation}
-\frac{e^{2\beta}}{R^2}+ \frac{ \alpha' R'}{R} + \frac{R^{'2}}{R^2} - e^{-2\alpha} e^{2\beta} \biggl(-\frac{2 \dot{R} \dot{\alpha}}{R} + \frac{2 \ddot{R}}{R^2} + \frac{\dot{R}^2}{R^2}\biggr) =  e^{-2\beta}(p_r - \frac{4}{3}\eta \sigma - \theta \zeta - \Lambda),
\end{equation}
Multiplying $G_{11} $ by $R^2 \dot{R}$ the above equation reduces to
\begin{equation}
[R(1+ \dot{R}^2 e^{-2\alpha} + R^{'2} e^{-2\beta} - (1/3)\Lambda R^2 )]_{,t}   = -(p_r - (4/3)\eta \sigma - \zeta \theta)R^2 \dot{R}.
\end{equation}
Comparing with the above equations with $F'$ and $\dot{F}$, we get the mass function as follows:
\begin{equation}
F = R\,[1+ \dot{R}^2 e^{-2\alpha} + R^{'2} e^{-2\beta} - (1/3)\Lambda R^{2}\, ].   
\end{equation}
These equations have been used in the main text.

\subsection{Integration of field equations: shell dynamics} \label{Appendix1}
In the following, we show the integration of the 
shell equation to obtain the dynamics of shells, $R = R(t)$.
\subsubsection{Marginally bound system}
Consider the equation of motion of the matter shells given in the main text in eqn. \eqref{eom} for $k(r)=0$: 
\begin{equation}
    \dot{R}(r,t)^2 = \frac{F(r)}{R} + \frac{1}{3} \Lambda R^{2} .
\end{equation}
Substituting $ n = (\Lambda/3)$, and $ F/n + R^{3} = Y$, with 
$ dR= dY/[3(Y-F/n)^{2/3}] $, leads to 
\begin{equation}
 t = \frac{1}{3\sqrt{n}} \int \frac{1}{\sqrt{Y} \sqrt{Y - F/n}} dY .
\end{equation}
Again, substitution of $ Y = (F/n) X $ $\implies$ $ dY = \frac{F}{n} dX $, and $X = 2 \cos^2{\theta} $ implies 
\begin{equation}
 t = -\frac{2 \sqrt{2}}{3 \sqrt{n}} \int \frac{\tan{\theta}}{\sqrt{1-\tan^2{\theta}}} d\theta .
\end{equation}
Using $ 1 - \tan^2{\theta} = 2Z^2 $, we get 
\begin{equation}
 t = \frac{2}{3 \sqrt{n}} \tanh^{-1}{(Z)} + C ,
\end{equation}
where $Z = \sqrt{\frac{1/3 \Lambda R^3}{F + 1/3 \Lambda R^3}}$.
This implies that 
\begin{equation*}
   t = -\frac{2}{\sqrt{3 \Lambda}} \tanh^{-1}\left({\sqrt{\frac{1/3 \Lambda R^3}{F+ 1/3 \Lambda R^3}}}\right) +t_{s}.
\end{equation*}
This equation may now be rewritten as 
\begin{equation}
 R = \left( \frac{3 F}{\Lambda}\right)^{1/3}  \sinh^{2/3} {\left[\frac{\sqrt{3 \Lambda}}{2}(t_s-t)\right]}  .
\end{equation}
By setting the singularity time $t_s$ = 0, the equation reduces to:
\begin{equation}
 R = \left( \frac{3 F}{\Lambda}\right)^{1/3}  \sinh^{2/3} {\left(\frac{\sqrt{3 \Lambda}}{2}t\right)} . 
\end{equation}
The constant of integration may be fixed as follows:
At $t = t_{i} = 0$, $R = r$, and therefore we get:
\begin{equation*}
     \int_{t_i = 0}^{t} dt = -\int_{r}^{R} f(R) \,dR,
\end{equation*}
where the negative sign is kept show that the process implies collapse of matter. Hence, we get:
\begin{equation*}
 t = -\frac{2}{\sqrt{3 \Lambda}} \tanh^{-1}\left({\sqrt{\frac{1/3 \Lambda R^3}{F+ 1/3 \Lambda R^3}}}\right) + \frac{2}{\sqrt{3 \Lambda}} \tanh^{-1}\left({\sqrt{\frac{1/3 \Lambda r^3}{F+ 1/3 \Lambda r^3}}}\right) .
\end{equation*}

\subsection{Gravitationally bound configurations}
\label{Appendix2}
The $k(r)>0$ signifies the bound collapse for which the velocity of the matter shells at the beginning of the collapse is negative.
The equation of motion of the collapsing shell \eqref{eom} for $k(r) \neq 0$ reduces to
\begin{equation}
\dot{R}(r,t)^2 = -k(r) + \frac{F(r)}{R} + \frac{1}{3} \Lambda R^2.
\end{equation}
The equation may be written in the following way:
\begin{eqnarray}
   t &=& -\int \Bigl[R^{-1} \Bigl(-k(r) R + F(r) + \frac{1}{3} \Lambda R^3 \Bigr) \Bigr]^{-1/2} dR, \nonumber \\
   &=& -\frac{(F(r)/4)^{1/3}}{(F(r)/4)^{1/3}} \int R^{1/2} \Bigl(-k(r) R + F(r) + \frac{1}{3} \Lambda R^3 \Bigr) ^{-1/2} dR .
\end{eqnarray}
Let us introduce an auxiliary variable:
\begin{equation}
   \eta = -\bigg(\frac{F(r)}{4}\bigg)^{1/3} \int \Bigl(-k(r) R^2 + F(r)R + \frac{1}{3} \Lambda R^4 \Bigr) ^{-1/2} dR .
\end{equation}
Defining new variable $X=-\frac{1}{R}$ $\implies$ $dR=\frac{1}{X^2} dX$
\begin{equation*}
  \eta  = \bigg(\frac{F(r)}{4}\bigg)^{1/3} \int \Bigl(-k(r) X^2 - F(r) X^3 + \frac{1}{3} \Lambda \Bigr)^{-1/2} dX .
\end{equation*}
Now, defining $X$ in the form of $Y$ such that
\begin{equation*}
    X = -\frac{Y - \delta(r)}{(F(r)/4)^{1/3}} \implies  dX = -\frac{dY}{(F(r)/4)^{1/3}},
\end{equation*}
where we have introduced $\delta(r)=\frac{1}{3}(-k)(2 F)^{-2/3}$.
The integration is well known and is given by:
\begin{equation*}
  \eta  = -\int [4 Y^{3} - g_2 Y - g_3]^{-1/2} dY ,
\end{equation*}
where $g_2 = 12 \delta(r)^{2}$ and $g_3 = -8\delta(r)^{3}-\frac{1}{3} \Lambda$.
Thus, if we define $Y = \mathcal{P}(\eta)$, and resubstitute, we get:
\begin{equation}
    R = -\frac{1}{X} = \frac{(F(r)/4)^{1/3}}{\mathcal{P}(\eta) - \delta(r)} .
\end{equation}
 Now, we determine the time coordinate in terms of the auxiliary variable $\eta$: 
\begin{eqnarray}
    t &=& \frac{1}{(F(r)/4)^{1/3}} \int R d\eta = \frac{1}{(F(r)/4)^{1/3}} \int \frac{(F(r)/4)^{1/3}}{\mathcal{P}(\eta) - \delta(r)} \nonumber\\
&= & \int \frac{1}{\mathcal{P}(\eta) - \mathcal{P}(\epsilon)} d\eta .
\end{eqnarray}

where $ \mathcal{P}(\epsilon) = \delta(r) $.
By multiplying and dividing the above equation by $ \mathcal{P'}(\epsilon) $, we get
\begin{equation}
  t =  \frac{1}{\mathcal{P'}(\epsilon)} \int \frac{\mathcal{P'}(\epsilon)}{\mathcal{P}(\eta) - \mathcal{P}(\epsilon)} d\eta  .
\end{equation}
Using: $\frac{\mathcal{P'}(\epsilon)}{\mathcal{P}(\eta) - \mathcal{P}(\epsilon)} = \zeta(\eta - \epsilon) - \zeta(\eta + \epsilon) + 2\zeta(\epsilon)$, we get
\begin{equation}
  t = \frac{1}{\mathcal{P'}(\epsilon)} \int \Bigl[\zeta(\eta - \epsilon) - \zeta(\eta + \epsilon) + 2\zeta(\epsilon) \Bigr] d\eta  .
\end{equation}
Using the fact that $ [\ln \sigma(z)]^{\prime} = \zeta (z) $, we rewrite:
\begin{eqnarray}
  t &=&  \frac{1}{\mathcal{P'}(\epsilon)} \Bigl[\ln \sigma(\eta - \epsilon) - \ln \sigma(\eta + \epsilon) + 2 \zeta(\epsilon) \eta \Bigr] + C    \nonumber\\
&=& \frac{1}{\mathcal{P'}(\epsilon)} \Bigl[\ln \Bigl(\frac{ \sigma(\eta - \epsilon)}{\sigma(\eta + \epsilon)}\Bigr) + 2 \zeta(\epsilon) \eta \Bigr] + C   .
\end{eqnarray}
Now, the invariants $g_{2}$  and $g_{3}$ of $\mathcal{P}(\epsilon)$ are not determined. These parameters must be chosen such that:
\begin{equation}
 \int \frac{1}{(4Z^3 -g_{2}Z-g_{3})^{-1/2}} dZ=\epsilon .
\end{equation}
This solves the equation.
\subsubsection*{Fluids admitting tangential pressure only}
For the fluid having only tangential pressure, $p_{r}=0$ and $\eta=\zeta=0$ and
assuming the equation of state of the form $p_{t}=k_{t}\rho$, the equation for collapsing shell is given by eqn. \eqref{tanp_shell_eqn}.
\begin{equation*}
\dot{R} = -R^{-1/2} \left[\frac{F(r)}{R} - 1 + \frac{k(r)}{R} + \frac{1}{3} \Lambda R^2 \right]^{1/2}    
\end{equation*}
This may be written as
\begin{equation*}
dt = -\frac{R}{\sqrt{I(r) -R + \frac{1}{3} \Lambda R^3}} dR  
\end{equation*}
%
%
where $F(r)+k(r)=I(r)$. This implies that the time curve is:
%
%
%
\begin{equation*}
dt = -\sqrt{\frac{12}{\Lambda}} \frac{R}{\sqrt{4 R^3 -\frac{12}{\Lambda} R + \frac{12}{\Lambda} I(r)}} dR  
\end{equation*}
Let us consider, just like before, another auxiliary variable:
\begin{eqnarray}
\eta &=& \int \frac{1}{\sqrt{4 R^3 -\frac{12}{\Lambda} R + \frac{12}{\Lambda} I(r)}} dR \nonumber\\
 &=& \int \frac{1}{\sqrt{4 R^3 - g_{2} R - g_{3}}} dR  
\end{eqnarray}
where $g_{2} = \frac{12}{\Lambda}$ and $g_{3} = -\frac{12}{\Lambda}I(r).$
Thus, $R = \mathcal{P}(\eta + \epsilon)$ where $\epsilon$ is a constant. Now, we get the solution of the time curve:
\begin{eqnarray}
t = -\sqrt{\frac{12}{\Lambda}} \int R d\eta \nonumber\\
t = -\sqrt{\frac{12}{\Lambda}} \int \mathcal{P}(\eta + \epsilon) d\eta\nonumber\\
t = +\sqrt{\frac{12}{\Lambda}} \zeta(\eta + \epsilon) + C
\end{eqnarray}
where C is the constant of integration.
%

\subsection{Junction Condition}
For the junction condition, we shall assume that the external spacetime is taken to be the static spherically symmetric  Schwarzschild- deSitter (Kottler) spacetime, given in \eqref{Kottler_metric}, whereas the internal spacetime is the 
metric in eqn. \eqref{guv}. The junction conditions are to be implemented on a $r=$ constant surface, where the metric and the extrinsic curvatures must be continuously joined. The details of the calculation is carried out in many references, and may be consulted \cite{banerjee, Lake:1978zz}. The matching leads in effect to the statement that, on the horizon, the mass function $F(r,t)$ must equal the quantity $2 M$ of the Kottler spacetime.  

%



\end{document}